\documentclass[twocolumn,pra, aps,superscriptaddress,floatfix]{revtex4}

\usepackage{lineno}
\usepackage{mathptmx}
\usepackage{subfigure}
\usepackage{dcolumn}
\usepackage{amsmath,amssymb}
\usepackage{bm}
\usepackage{color}
\usepackage{overpic}
\usepackage{latexsym}
\usepackage{epstopdf}
\usepackage{color}
\usepackage[english]{babel}
\usepackage{latexsym}
\usepackage{stmaryrd}

\usepackage{psfrag,graphicx} 
\usepackage{epsf} 
\usepackage{subfigure} 
\usepackage{amsmath} 
\usepackage{amssymb} 
\usepackage{amsfonts}
\usepackage{bm}
\usepackage{natbib}
\usepackage{epstopdf}
\usepackage{appendix}

\definecolor{mygrey}{gray}{0.35}
\definecolor{myblue}{rgb}{0.2,0.2,0.8}
\definecolor{myzard}{cmyk}{0,0,0.05,0}
\definecolor{mywhite}{rgb}{1,1,1}
\definecolor{myred}{rgb}{1,0.,0.3}

\usepackage[colorlinks=true,citecolor=myblue,linkcolor=myred]{hyperref}
\def\be{\begin{equation}}
\def\ee{\end{equation}}
\def\ba{\begin{align}}
\def\enda{\end{align}}
\def\bi{\begin{itemize}}
\def\ei{\end{itemize}}

\def\dd{\mathord{\rm d}} 
 \def\ee{\mathord{\rm e}}
 
 \def\ii{\mathord{\rm i}}

\def\half{\textstyle\frac{1}{2}}

\def\fourth{\textstyle\frac{1}{4}}

\def\dd{\mathord{\rm d}} 
 \def\ee{\mathord{\rm e}}
 
 \def\ii{\mathord{\rm i}}

\def\half{\textstyle\frac{1}{2}}

\def\fourth{\textstyle\frac{1}{4}}

\renewcommand{\ii}{{\rm i}}
\renewcommand{\ee}{{\rm e}}

\def\beq{\begin{equation}}
\def\beq{\begin{equation}}
\def\eeq{\end{equation}}

 \newcommand{\ket}[1]{|#1\rangle}
 \newcommand{\bra}[1]{\langle #1|}

\begin{document}


\title[Short Title]{Exploring Interacting Topological Insulators with Ultracold Atoms: \\ the Synthetic Creutz-Hubbard Model}

\author{ J. J\"{u}nemann}
\affiliation{Johannes Gutenberg-Universit\"{a}t, Institut f\"{u}r Physik, Staudingerweg 7, 55099 Mainz, Germany }
\affiliation{MAINZ - Graduate School Materials Science in Mainz, Staudingerweg 9, 55099 Mainz, Germany}

\author{A. Piga}
\affiliation{ICFO-Institut de Ciencies Fotoniques, The Barcelona Institute of Science and Technology, 08860 Castelldefels (Barcelona), Spain}

\author{S.-J. Ran}
\affiliation{ICFO-Institut de Ciencies Fotoniques, The Barcelona Institute of Science and Technology, 08860 Castelldefels (Barcelona), Spain}

\author{M. Lewenstein}
\affiliation{ICFO-Institut de Ciencies Fotoniques, The Barcelona Institute of Science and Technology, 08860 Castelldefels (Barcelona), Spain}
\affiliation{ICREA, Lluis Companys 23, 08010 Barcelona, Spain}

\author{M. Rizzi}
\affiliation{Johannes Gutenberg-Universit\"{a}t, Institut f\"{u}r Physik, Staudingerweg 7, 55099 Mainz, Germany }

\author{A. Bermudez}
\affiliation{Department of Physics, Swansea University, Singleton Park, Swansea SA2 8PP, United Kingdom}
\affiliation{Instituto de F\'{i}sica Fundamental, IFF-CSIC, Madrid E-28006, Spain}

\begin{abstract}
Understanding the robustness of topological phases of matter in the presence of strong interactions, and synthesising novel strongly-correlated topological materials, lie among the most important and difficult challenges of modern theoretical and experimental physics. In this work, we present a complete theoretical analysis of the synthetic Creutz-Hubbard ladder, which is a paradigmatic model that provides a neat playground to address these challenges. We put special attention to the competition of correlated topological phases and orbital quantum magnetism in the regime of strong interactions. These results are furthermore confirmed and extended by extensive numerical simulations. Moreover we propose how to experimentally realize this model in a synthetic ladder, made of two internal states of ultracold fermionic atoms in a one-dimensional optical lattice. Our work paves the way towards quantum simulators of interacting topological insulators with cold atoms.

\end{abstract}

\maketitle
\begingroup
\hypersetup{linkcolor=black}
\tableofcontents
\endgroup

\section{Introduction}
\label{sec:introduction}

Topological features of quantum many-body systems provide a new paradigm in our understanding of the phases of matter~\cite{wen_book}, and give rise to a promising avenue towards fault-tolerant quantum computation~\cite{top_quantum_codes}. From a condensed-matter perspective, such features lead to exotic groundstates beyond the conventional phases of matter, which are typically understood by the principle of symmetry breaking and the notion of a local order parameter. On the contrary, these exotic states can only be characterised by certain topological properties. 

The integer quantum Hall effect, which is a paradigmatic example of such peculiar phases~\cite{iqhe}, requires the introduction of a topological invariant to describe the different plateaus and their associated transverse conductivities~\cite{tkkn}. Another interesting property of this state of matter is the bulk-boundary correspondence, which relates such a topological conductivity, a bulk property, to the existence of current-carrying edge states localised within the boundaries of the system~\cite{edge_states_iqhie}. Although the bulk of an integer quantum Hall sample appears as a trivial band insulator, its 
boundary corresponds to a holographic chiral liquid where interactions merely renormalise the edge-state Fermi velocity~\cite{edge_transport}.

As realised in a series of seminal works~\cite{haldane_model,kitaev_model,kane_mele}, these remarkable properties are not unique to quantum Hall samples subjected to strong magnetic fields. Instead, they arise in various models with different symmetries and in different dimensions~\cite{top_ins_table}, the so-called topological insulators and superconductors~\cite{top_ins_review}, which also lead to the notion of \emph{symmetry-protected topological phases} in the context of topological order~\cite{wen_book}.
Remarkably enough, some of these models have turned out to be accurate descriptions of real insulating materials~\cite{2d_qsh_exp,3d_ti_exp,top_ins_review}, and promising candidates to account for observations in proximitized superconducting materials~\cite{majoranas_exp}. This has positioned the subject of topological insulators and superconductors not only at the forefront of academic research, but also at the focus of technological applications, such as topological quantum computation with Majorana fermions~\cite{top_quantum_codes}. 

Despite this success, \emph{(i)} there are still several paradigmatic models of topological models whose connection to real materials still remains unknown, or even seems quite unlikely, as it occurs for the Hofstadter model with magnetic fluxes on the order of the flux quantum~\cite{hofstadter}, or the Haldane model~\cite{haldane_model}. Moreover, \emph{(ii)} most of the topological materials explored in the laboratory so far do not display important electronic \emph{correlation effects}~\cite{interacting_ti_review}. This is rather unfortunate in view of the richness of the fractional quantum Hall effect~\cite{fractional_qhe}, where such correlations are responsible for a plethora of exotic topological phases of matter.

In the present work, rather than considering real materials, we shall be concerned with the so-called \emph{synthetic quantum matter} in  atomic, molecular and optical (AMO) platforms, more particularly with ultracold gases of neutral atoms trapped in periodic potentials made of light, i.e. optical lattices~\cite{review_cold_atoms}.
The ever-improving experimental control over these quantum many-body systems has already allowed to design their microscopic Hamiltonian to a great extent.
In this way, it is possible to target interesting condensed-matter models, some of which are still lacking unambiguous experimental feedback from experiments with real materials~\cite{lewenstein_review,qs_cold_atoms}, as occurs for the bosonic~\cite{bose_hubbard,bose_hubbard_ol,bhm_ol_exp} and fermionic~\cite{hubbard,fermi_hubbard_ol,fhm_ol_exp} Hubbard models.
The initial interest in optical-lattice implementations of integer quantum Hall phases~\cite{jz_gauge_fields}, and other time-reversal invariant topological insulators~\cite{tr_hofstadter,wilson_fermions}, has risen considerably in the recent years due to the experimental progress~\cite{goldman_review,goldman_news_views}.
In particular, elusive topological systems such as the Hofstadter~\cite{hofstadter} or the Haldane~\cite{haldane_model} models, have been realised in experiments with bosonic~\cite{gauge_ol} and fermionic~\cite{haldane_ol} ultracold atoms, respectively.
The possibility of including two or more different atomic states/species, controlling their interactions via Feshbach resonances~\cite{feshbach}, may eventually lead to the experimental test of correlation effects in these topological states.

\begin{figure}
\centering
\includegraphics[width=0.8\columnwidth]{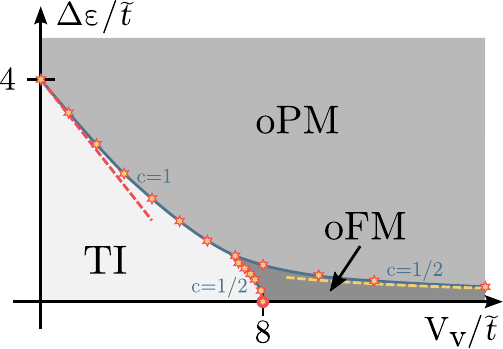}
\caption{ {\bf Phase diagram of the imbalanced Creutz-Hubbard ladder: } Phase diagram displaying a topological insulator (TI) phase, and a pair of non-topological phases: an orbital phase with long-range ferromagnetic Ising order (oFM), and an orbital paramagnetic phase (oPM). The horizontal axis represents the ratio of the inter-particle interactions to the tunneling strength, whereas the vertical axis corresponds to the ratio of the energy imbalance to the tunneling strength. The dashed yellow line shows the transition points of the effective model in the strong-coupling effective (Ising) model. The dashed red line indicates the transition as obtained from the weak-coupling expansion. The red circle shows the transition point in the balanced model at intermediate interactions. Stars label numerical results, and the blue line is an extrapolation of the phase-boundaries. The labels of the critical lines give the central charge of their underlying conformal field theory.
Details on the different phases and transitions between them are provided in Sec.~\ref{sec:int_tqpt}.}
\label{fig_phase_diagram}
\end{figure}

Here, we introduce a variant of the quasi-one-dimensional (quasi-1D) {Creutz topological insulator}~\cite{creutz_ladder}, and study the effect  of repulsive  Hubbard-type interactions on the topological phase. This model, which shall be referred to as the {\it imbalanced Creutz-Hubbard ladder},  is readily implementable with ultracold fermions in intensity-modulated optical lattices. We argue that such a  model has all the required ingredients to become a workhorse in the study of strongly-correlated topological phases in AMO setups.
Indeed, two accessible AMO ingredients as {\it (i) }a simple \emph{Zeeman shift} between the atomic internal states, which yields a leg imbalance in the ladder, and {\it (ii)} \emph{on-site repulsive contact interactions} tuned by some \emph{Feshbach resonances}, which lead to Hubbard-type rung interactions in the ladder, can be employed to access the rich phase diagram of this model (see Fig.~\ref{fig_phase_diagram}), which  was widely uncharted prior to our  study. 
Starting from a \emph{flat-band} regime, we show that the imbalance and the interactions lead to a competition  between a topological phase and two different phases of orbital quantum magnetism.
At large interaction strength, a long-range in-plane ferromagnetic order arises, related to the symmetry-broken phase of an orbital quantum Ising model; while the Zeeman imbalance then drives a standard quantum phase transition in the Ising universality class towards an orbital paramagnetic phase.
In order to understand the model away from this limit, we introduce two new methods based on mappings onto models of quantum magnetism and quantum impurity physics. 
These methods allow us to locate exactly certain critical points/lines, and to predict topological quantum phase transitions for weak and intermediate interactions with different underlying conformal field theories (CFTs), i.e. Dirac versus Majorana CFTs, which then fit very well with numerical results based on Matrix-Product-States (MPS)~\cite{mps_review}. 
We also provide suggestions for experimental observables to pinpoint these three phases.

This manuscript is organized as follows: in Sec.~\ref{sec:chl} the standard Creutz ladder is introduced,  
and some previous  studies are briefly accounted for, which show that this model leads to a topological insulator in the BDI symmetry class~\cite{top_ins_table}. Therefore, the standard Creutz ladder lies in the same symmetry class as the Su-Schrieffer-Heeger model~\cite{ssh_model}, which has already been implemented in optical lattices~\cite{measurement_zak_phase_ssh}. 
We then introduce the imbalanced Creutz ladder, which provides an instance of a topological insulator in the AIII class (chiral unitary), which 
still lacks an AMO implementation. Moreover, as argued in Sec.~\ref{sec:int_tqpt}, the Hubbard interactions lead to a very neat interplay of strongly-correlated and topological effects.  We identify the different phases of the model, together with standard and topological quantum phase transitions that connect them. Finally, in Sec.~\ref{sec:proposal}, we lay out a detailed proposal to implement the imbalanced Creutz-Hubbard ladder in an ultracold Fermi gas with two different internal states trapped in a standard one-dimensional (1D) optical lattice, enriched by intensity-shaking and Raman laser-assisted tunnelling. We present our conclusions and outlook in Sec.~\ref{sec:conclusions}.


\section{The imbalanced Creutz-Hubbard ladder}
\label{sec:chl}

\begin{figure}
\centering
\includegraphics[width=0.85\columnwidth]{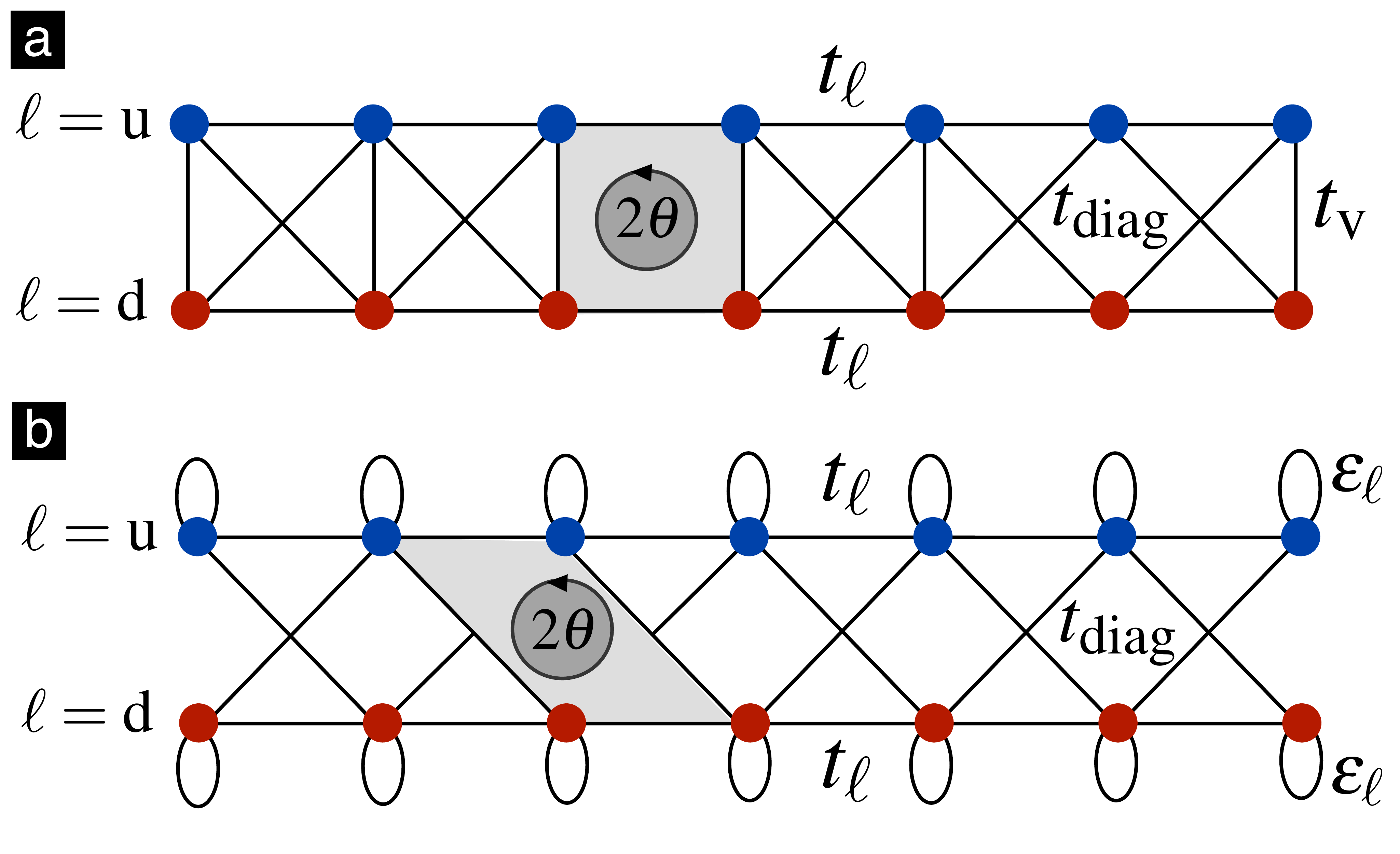}
\caption{ {\bf Standard and imbalanced Creutz ladder: } Two-leg ladder where fermions tunnel along the black links enclosing a net flux $2\theta$ along a closed plaquete: {\bf (a)} standard Creutz  ladder~\eqref{or_cl}, and  {\bf (b)} imbalanced Creutz  ladder, which leads to Eq.~\eqref{eq:creutz-hubbard_flat} in the $\pi$-flux limit.}
\label{fig_cl_scheme}
\end{figure}

The standard Creutz model describes a system of spinless fermions on a two-leg ladder (see Fig.~\ref{fig_cl_scheme} {\bf (a)}), which are created-annihilated by the fermionic operators $c_{i,\ell}^{\dagger},c_{i,\ell}^{\phantom{\dagger}}$, where    $i\in\{1,\dots,N\}$ labels the lattice sites within each leg of the ladder $\ell\in\{\rm u,d\}$. Fermions are allowed to hop vertically along the rungs of the ladder with tunnelling strength $t_{\rm v}$, and horizontally along
the legs of the ladder with a complex tunnelling $t_{\ell}=t_{\rm h}(\ee^{\ii\theta} \delta_{\ell,\rm u}+\ee^{-\ii\theta}\delta_{\ell,\rm d})$, where $t_{\rm h}$ is a tunnelling strength, and $\delta_{a,b}$ is the Kronecker delta. The arrangement of complex phases in the horizontal links leads to a net 2$\theta$-flux gained by a fermion hopping around a square unit cell, playing thus the role of the so-called Peierls phases of a magnetic field piercing the ladder. In addition, the kinetic part of the Hamiltonian also includes a diagonal tunnelling of strength $t_{\rm diag}$, yielding altogether 
\beq
\label{or_cl}
H_{\rm C}=-\sum_{i}\sum_{\ell}\left(t_{\ell}c_{i+1,\ell}^{\dagger}c_{i,\ell}^{\phantom{\dagger}}+t_{\rm diag}c_{i+1,\ell}^{\dagger}c_{i,\bar{\ell}}^{\phantom{\dagger}}+t_{\rm v}c_{i,\ell}^{\dagger}c_{i,\bar{\ell}}^{\phantom{\dagger}}+{\rm H.c.}\right),
\eeq
where we use the notation $\bar{\ell}= {\rm d} (\bar{\ell}= {\rm u}) $ for $\ell= {\rm u}$ ($\ell= {\rm d}$).

This quadratic lattice model was put forth in Ref.~\cite{creutz_ladder} as a simple toy model to understand  some of the key properties of higher-dimensional domain-wall fermions~\cite{kaplan_fermions,hamiltonian_kaplan_fermions}, which were introduced in the context of lattice gauge theories to bypass the fermion-doubling problem~\cite{fermion_doubling}. For periodic boundary conditions, this model leads to a couple of bands that display a pair of massive Dirac fermions with different Wilson masses $m_{0},m_\pi$ at momenta $k_{\rm D}\in\{0,\pi\}$~\cite{wilson_fermions_b}. 
For open boundary conditions, a pair of zero-energy modes exponentially localised to the left/right edges of the ladder appear as one of the Wilson masses gets inverted ($m_\pi<0$) when $t_{\rm v}<2t_{\rm diag}$. Considering the bulk-boundary correspondence discussed in the introduction, these edge states resemble the holographic liquid of the higher-dimensional topological insulators. In fact, the change in polarisation of the system can be characterised by a topological invariant~\cite{berry_phase_effects}, the so-called Zak's phase~\cite{zak}, such that the appearance of these zero-energy modes coincides with a non-vanishing topological invariant, and the Creutz ladder yields a symmetry-protected topological phase in this regime. As discussed below, for $\theta=\pi/2$, this topological phase corresponds to a BDI topological insulator.

Since the objective of this work is to study correlation effects, we now consider the simplest possible Hubbard interactions between the spinless fermions
\beq
\label{eq:H_H}
H_{\rm H}=\sum_{i}\sum_\ell \left(V_{\rm h}n_{i,\ell}n_{i+1,\ell}+\frac{V_{\rm v}}{2}n_{i,\ell}n_{i,\bar{\ell}}\right),
\eeq
where $V_{\rm h}$ ($V_{\rm v}$) are the density-density interaction strengths between fermions residing in neighbouring sites along horizontal (vertical) bonds of the ladder, and we have introduced the fermion number operators $n_{i,\ell}=c^\dagger_{i,\ell}c_{i,\ell}^{\phantom{\dagger}}$.

For reasons that will become clear in the cold-atom implementation discussed in Sec.~\ref{sec:proposal}, in the following we will deal with a variant of the Creutz Hamiltonian:
\emph{(i)} we shall substitute the vertical tunnelling by an energy imbalance between the legs of the ladder $\epsilon_{\rm u}=\Delta\epsilon/2=-\epsilon_{\rm d}$, which changes the symmetry class of  the  topological insulator for $\theta=\pi/2$ from BDI to AIII;
\emph{(ii)} we limit the interaction terms~\eqref{eq:H_H} to the anisotropic regime $V_{\rm h}=0$;
\emph{(iii)} we set the amplitude of the diagonal hopping equal to the one along the legs ($|t_{\rm diag}| = |t_\ell|=\tilde{t}$) and finally %
\emph{(iv)} we fix the phases in order to get a net $\pi$-flux through the plaquettes.
The resulting Hamiltonian (see Fig.~\ref{fig_cl_scheme} {\bf (b)}), which we will refer to is the \emph{imbalanced Creutz-Hubbard Hamiltonian}, is
\beq 
\label{eq:pi_ch}
H_{\rm \pi CH}=H_{\rm \pi C}+V_{\rm Hubb},\hspace{2ex} H_{\rm \pi C}=H_{\rm FB}+V_{\rm imb},
\eeq 
where we have introduced the kinetic term
\beq
\label{eq:creutz-hubbard_flat}
H_{\rm FB}=\sum_{i,\ell}\left(-\tilde{t}c_{i+1,\ell}^{\dagger}c_{i,\bar{\ell}}^{\phantom{\dagger}}+\ii s_\ell \tilde{t}c_{i+1,\ell}^{\dagger}c_{i,{\ell}}^{\phantom{\dagger}}+{\rm H.c.}\right)
\eeq
with $s_{u/d}=\pm1$. This term leads to a pair of flat bands, and a couple of zero-energy topological edge states. The remaining terms 
\beq
\label{eq:creutz-hubbard_pert}
V_{\rm imb}=\sum_{i,\ell} \frac{\Delta\epsilon}{2} s_\ell n_{i,{\ell}}^{\phantom{\dagger}},\hspace{2ex} V_{\rm Hubb} = \sum_{i,\ell} \frac{V_{\rm v}}{2}n_{i,\ell}n_{i,\bar{\ell}},
\eeq
contain the Hubbard interactions and the energy imbalance, which can have non-trivial effects on the flat-band physics, and induce a phase transition to other non-topological phases of matter. In the following section, we present a new formalism to understand such transitions.

We are encouraged to study this {imbalanced Creutz-Hubbard ladder} both for experimental and theoretical reasons that will be discussed in detail in the forthcoming sections. In particular, let us advance that this model shares topological properties with the original Hamiltonian~\eqref{or_cl}. In particular, the Dirac fermions now occur at $k_{\rm D}\in\{-\pi/2,\pi/2\}$ with different Wilson masses $m_{\pm\pi/2}$. Provided that $\Delta\epsilon<4t_{\rm h}=4t_{\rm d}$, the mass $m_{\pi/2}<0$ gets inverted, and we obtain analogous topological features and exponentially-localised edge states. The choice of the anisotropic regime $V_{\rm h}=0$ is motivated by the use of ultracold fermionic atoms in optical lattices with contact interactions. Exploring also the regime of $V_{\rm h}>0$ would require situations where the atoms have longer-range interactions.

\subsection*{Previous studies on related models}\label{ssec:previousstudies}

With the model under study been defined, let us comment on some relevant literature, and advance some comments in relation to our results.  
The standard Creutz ladder~\eqref{or_cl} filled with bosons has been studied in Refs.~\cite{bosonic_creutz_hubbard_aoki,bosonic_creutz_hubbard_huber}, which includes on-site Hubbard interactions instead of the nearest-neighbour terms of Eq.~\eqref{eq:H_H}. The focus of these two articles was the appearance of pair superfluids due to the interplay of interactions and frustration flat-band effects~\cite{bose_flat_bands_altman}. A useful formalism was introduced in these works, which consisted on the projection of the Hamiltonian onto the lowest-energy flat band by introducing highly-localised Wannier functions. Moreover, using a unitarily-equivalent formulation of the Creutz-ladder Hamiltonian introduced in Ref.~\cite{bosonic_creutz_hubbard_aoki} would lead  to a spin-orbit coupled Hubbard model with staggered energy imbalance in our case~\eqref{eq:pi_ch}, which might broaden the relevance of our results beyond the ladder compound of Fig.~\ref{fig_cl_scheme}.
The standard Creutz ladder~\eqref{or_cl} populated by spinful fermions with on-site interactions between opposite spins has been studied recently in Ref.~\cite{creutz_Hubbard}, which focused on an exact Bardeen-Cooper-Schrieffer description for attractive Hubbard interactions.
To the best of our knowledge, the standard Creutz ladder  with spinless fermions has only been studied previously in Ref.~\cite{creutz_majorana_hubbard}. Here, the emphasis was placed on an additional superconducting $s$-wave pairing, and its interplay with the repulsive Hubbard interactions, which leads to an interesting Creutz-Majorana-Hubbard ladder. The limit of vanishing pairing, which corresponds to the standard Creutz-Hubbard model with the vertical tunneling~\eqref{or_cl} instead of the imbalance~\eqref{eq:creutz-hubbard_pert}, and leads to the aforementioned BDI topological insulator, was only touched upon briefly by numerical mean-field and density-matrix renormalisation group studies. These results pointed towards the possibility of a phase with in-plane ferromagnetic order as the interactions are increased.

To the best of our knowledge, the effect of Hubbard interactions in the imbalanced Creutz ladder~\eqref{eq:pi_ch}, and more generally on AIII topological insulators, remains largely unexplored. In relation to the above  studies on related models, we show below that one cannot project onto a single flat band for the fermionic model, but must instead retain all flat bands and edge modes to account for  correlation and topological effects accurately. In this way, we shall make interesting connections between the existence of topological phase transitions and  the physics of quantum impurity models. Additionally, by going beyond a mean-field analysis in the strong-interaction limit of the imbalanced model~\eqref{eq:pi_ch}, we show analytically that in-plane ferromagnetic order also arises in our model, and moreover  corresponds to the symmetry-broken phase of an orbital quantum Ising model. In addition, we illustrate that the imbalance drives such Ising model into a orbital paramagnetic phase. 


\section{Topological quantum phase transitions}
\label{sec:int_tqpt}

The following subsections are devoted to the construction of the phase diagram shown in Fig.~\ref{fig_phase_diagram}.
We start by discussing the solution of the non-interacting imbalanced Creutz ladder, and the appearance of flat bands and fully-localised edge states in the Hamiltonian~\eqref{eq:creutz-hubbard_flat} (see Sec.~\ref{sec:flat_bands}). This corresponds to the vertical axis of the phase diagram. In Sec.~\ref{sec:weak_interactions}, we examine the weakly-interacting regime and show that the model maps onto a pair of weakly-coupled Ising chains, which can be studied through a mean-field analysis (i.e. the region in the vicinity of the vertical axis of Fig.~\ref{fig_phase_diagram}). In Sec.~\ref{sec:XYmodel}, we study the opposite limit of very strong interactions (i.e. rightmost region of  Fig.~\ref{fig_phase_diagram}), and discuss the possible non-topological orbital magnetic phases that can arise. In Sec.~\ref{sec:int_regime}, we explore the intermediate regime, and show that the effect of the interactions and imbalance leads to edge-bulk couplings that can be mapped onto quantum impurity models. This new perspective yields a neat picture underlying the destabilisation of the topological phase in 
favour of the orbital magnets. These different methods allow us to build an analytical prediction of the phase diagram of the model. Finally, in Sec.~\ref{sec:MPS}, we test numerically the above predictions, and provide a detailed study of the phase diagram by means of Matrix-Product-State numerical simulations. Scattered through these sections, we shall also introduce comments on possible experimental tools to measure the relevant observables for the different phases of the model, which will become relevant for the specific experimental cold-atom proposal discussed in Sec.~\ref{sec:proposal}.

\subsection{Non-interacting limit: Flat bands and edge states}
\label{sec:flat_bands}

We start by solving the kinetic part~\eqref{eq:creutz-hubbard_flat} of the $\pi$-flux Creutz-Hubbard Hamiltonian~\eqref{eq:pi_ch}. 
For periodic boundary conditions, and after introducing the spinor $\Psi(q)=(c_{\rm u}(q),c_{\rm d}(q))^{\rm t}$ for the fermion operators in momentum space $c_{\ell}(q)=\sum_i\ee^{-\ii qa i}c_{i,\ell}/\sqrt{N}$, one finds 
\beq\label{eq:FB_mom}
H_{\rm FB}=\sum_{q\in{\rm BZ}}\Psi^\dagger(q)\boldsymbol{B}(q)\cdot\boldsymbol{\sigma}\hspace{0.2ex}\Psi(q),
\eeq
where $\boldsymbol{\sigma}=(\sigma_x,\sigma_y,\sigma_z)$ is the vector of Pauli matrices, and $\boldsymbol{B}(q)=2\tilde{t}(-\cos (qa),0,\sin(qa))$. By direct diagonalization, 
one finds that the system develops two \emph{flat bands} $\epsilon_{\pm}:=\epsilon_{\pm}(q)=\pm 2\tilde{t}$, where $q\in{\rm BZ}=[-\pi/a,\pi/a)$ is the quasi-momentum, and the lattice constant shall be set to $a=1$ henceforth. The vanishing group velocity associated with these bands, $v_{\rm g}=\partial_q\epsilon_{\pm}(q)=0,$ indicates that the groundstate must be insulating, regardless of the particular filling. This can be considered as a new type of insulator, namely a \emph{flat-band insulator}, which     corresponds neither to the usual band insulator, nor to the Mott insulators. It shares some properties with the former (i.e. no correlations), and with the latter (i.e. localized fermions), but it differs from both insulators in the large degeneracy of the ground state, except for half-filling conditions.

On top of this, the flat bands are also \emph{topological}:
the diagonalization of the discrete chiral symmetry $\sigma_y$ (s.t. 
$\sigma_y H(q) \sigma_y = - H(q)$, with $H(q) = \boldsymbol{B}(q)\cdot\boldsymbol{\sigma}$)
puts the Hamiltonian \eqref{eq:FB_mom} in a purely off-diagonal form, with elements $B_x \pm i B_z$;
its complex phase gets a non-trivial winding number $\mathcal{W} = \mathrm{sgn}(\tilde{t}) \neq 0$~\cite{mazza_winding}.
Equivalently, we could consider the eigenvectors 
$\ket{\epsilon_{\pm}(q)} \propto \left( (B_x + i B_z)^{1/2} , \pm (B_x - i B_z)^{1/2} \right)^t$
and realize that they exhibit a uniform Berry connection 
$\mathcal{A}_\pm(q)=\ii\langle\epsilon_{\pm}(q)|\partial_q|\epsilon_{\pm}(q)\rangle=\half$.
The uniform Berry connection leads to a finite Zak's phase, which is defined~\cite{zak} as 
\beq
\label{eq:zak_phase}
\varphi_{\rm Zak,\pm}=\int_{\rm BZ}{\rm d}q\mathcal{A}_\pm(q) 
\eeq
and equals $\varphi_{\rm Zak,\pm}=\pi$ for our topological flat bands.
This Zak's phase pinpoints the topological properties of the bands, and can be connected to a macroscopic observable:
the polarization of the system~\cite{berry_phase_effects}.

Interestingly enough, the Creutz ladder displays an infinite flatness parameter without requiring long-range tunnelings, as necessary in higher-dimensional models of topological flat bands~\cite{topological_falt bands}.
From this perspective, switching on the leg imbalance $\Delta\epsilon>0$ in Eq.~\eqref{eq:creutz-hubbard_pert} leads to some curvature in the energy bands 
 \beq
 \label{eq:bulk_bands}
 \epsilon_{\pm}(q)=\pm\epsilon(q)=\pm2\tilde{t}\sqrt{1+\frak{f}^{-2}+2\frak{f}^{-1}\sin q},
 \eeq 
 where we have introduced the flatness parameter $\frak f=4\tilde{t}/\Delta\epsilon$, which becomes infinite for vanishing imbalance. 
The presence of the imbalance also drives the system out of the BDI class and into the AIII class~\cite{top_ins_table},
 since $B_z(q) =  \Delta \varepsilon / 2+ 2 \tilde{t} \sin q$ acquires a mixed parity under $k \leftrightarrow - k$ and therefore no effective time-reversal ($U_T H(-q)^* U_T^\dagger = + H(q)$) or particle-hole ($U_C H(-q)^* U_C^\dagger = - H(q)$) operators can be found.
Anyway, the discrete chiral symmetry is still described by $\sigma_y$ and the whole procedure described above can be employed.
The Berry connection of the bands becomes non-uniform due to their curvature
\beq
\mathcal{A}_\pm(q)=\frac{1+\frak{f}^{-1}\sin q}{2(1+\frak{f}^{-2}+2\frak{f}^{-1}\sin q)},
\eeq
which leads to the following Zak's phase $\varphi_{\rm Zak,\pm}=\pi\theta(\frak{f}-1)$, where $\theta(x)$ is the Heaviside step function. Hence, the Zak's phase yields topological effects provided that the bands are sufficiently flat, i.e. $\frak{f}>1$. Conversely, when $\Delta\epsilon>4\tilde{t}$, the curvature of the bands is large, i.e. $\frak{f}<1$, and no topological phenomenon occurs. 
We note that this point $\Delta\epsilon=4\tilde{t}$ is exactly the regime where one of the fermionic Wilson masses vanishes, as mentioned in the previous section in connection to the domain-wall fermions of lattice gauge theories. This marks a quantum phase transition between the AIII topological insulator and a trivial band insulator, as shown in the vertical axis of  Fig.~\ref{fig_phase_diagram}.

Regarding cold-atom experiments, we note that the Zak's phase has been measured for another paradigmatic topological insulator in 1D: the Su-Schrieffer-Heeger model of polyacetilene~\cite{ssh_model}. In this non-interacting limit, the Zak's phase is a single-particle property of the topological bands, and can be thus measured by a Ramsey interferometric protocol that exploits Bloch oscillations of single-particle initial states~\cite{measurement_zak_phase_ssh}. This measurement scheme has been generalised to other topological insulators~\cite{zak_phase_other_models}, and can also be applied to the cold-atom implementation of the Creutz ladder considered in Sec.~\ref{sec:proposal}.

\begin{figure*}
\centering
\includegraphics[width=1.7\columnwidth]{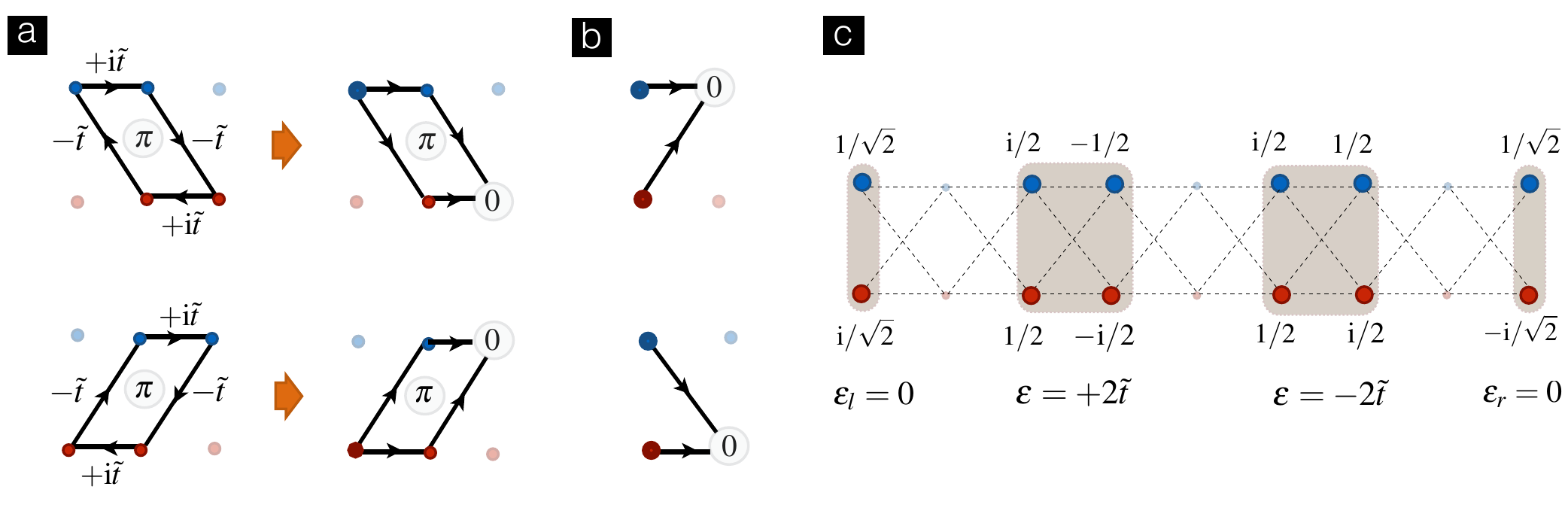}
\caption{ {\bf Aharonov-Bohm cages in the Creutz ladder: } {\bf (a)} Considering the tunnelling paths in the Creutz ladder~\eqref{eq:creutz-hubbard_flat}, one can identify two type of rhombic plaquettes that enclose a synthetic $\pi$-flux (left). Therefore, a particle trying to tunnel two sites apart (middle) will be subjected to a destructive Aharonov-Bohm interference that forbids this process. Accordingly, particles are confined to the so-called Aharonov-Bohm cages and cannot spread through the entire lattice. These cages correspond to square plaquettes, except at the edges {\bf (b)}, where destructive interference can also be found for a particle trying to tunnel one site apart (right). {\bf (c)} Aharonov Bohm cages with the relative amplitudes for a single-particle state in two possible flat bands, and in the two possible zero-energy edge modes.}
\label{fig_ab_cage_scheme}
\end{figure*}

To have an alternative view on these topological features, let us introduce the so-called { Aharonov-Bohm cages}~\cite{aharonov_bohm_cages}, which will also become very useful once interactions are switched on. In the $\pi$-flux Creutz ladder~\eqref{eq:creutz-hubbard_flat}, the fermions cannot tunnel two sites apart due to the Aharonov-Bohm effect~\cite{creutz_ladder} (see Fig.~\ref{fig_ab_cage_scheme}{\bf (a)}). One can thus find single-particle eigenstates strictly localised in cages formed by simple square plaquettes (see Fig.~\ref{fig_ab_cage_scheme}{\bf (c)}). In second quantisation, such Aharonov-Bohm cages (AB-c) with energies $\epsilon_{\pm}=\pm 2\tilde{t}$ are
\beq
\label{ab_cages}
\begin{split}
\ket{+2\tilde{t}}_i&=w_{i,+}^\dagger\ket{0},\hspace{1ex} w_{i,+}^\dagger=\frac{1}{2}\left(\ii c_{i,{\rm u}}^\dagger+c_{i,{\rm d}}^\dagger-c_{i+1,{\rm u}}^\dagger-\ii c_{i+1,{\rm d}}^\dagger\right),\\
\ket{-2\tilde{t}}_i&=w_{i,-}^\dagger\ket{0},\hspace{1ex} w_{i,-}^\dagger=\frac{1}{2}\left(\ii c_{i,{\rm u}}^\dagger+c_{i,{\rm d}}^\dagger+c_{i+1,{\rm u}}^\dagger+ \ii c_{i+1,{\rm d}}^\dagger\right),\\
\end{split}
\eeq
with $i\in\{1,\dots N\}$ for periodic boundary conditions, where one identifies $c_{N+1,\ell}=c_{1,\ell}$. Conversely, for open boundary conditions, these square AB-c can only be defined for $i\in\{1,\dots, N-1\}$, and simple counting shows that there are only $2(N-1)$ possible states that can be accommodated in such flat bands. The two missing states are zero-energy modes, $\epsilon_l=\epsilon_r=0$, fully localised at the boundaries 
\beq
\label{edge_cages}
\begin{split}
\ket{0}_{\rm L}&=l^\dagger\ket{0},\hspace{1.7ex} l^\dagger=\frac{1}{\sqrt{2}}\left( c_{1,{\rm u}}^\dagger+\ii c_{1,{\rm d}}^\dagger\right),\\
\ket{0}_{\rm R}&=r^\dagger\ket{0},\hspace{1ex} r^\dagger=\frac{1}{\sqrt{2}}\left( c_{N,{\rm u}}^\dagger-\ii c_{N,{\rm d}}^\dagger\right).\\
\end{split}
\eeq
For these particular weights, the fermions cannot tunnel one site apart due to the Aharonov-Bohm effect (see Fig.~\ref{fig_ab_cage_scheme}{\bf (a)}), and are thus localised within a boundary AB-c (see Fig.~\ref{fig_ab_cage_scheme}{\bf (b)}), which corresponds to an edge state within the bulk-edge correspondence of the topological insulator.

We can finally express the $\pi$-flux Creutz Hamiltonian~\eqref{eq:pi_ch} for open boundary conditions as
\beq
\label{eq:FB_2q}
H_{\rm FB}=\sum_{i=1}^{N-1}\sum_{\alpha=\pm}\epsilon_\alpha w_{i,\alpha}^\dagger w_{i,\alpha}^{\phantom{\dagger}}+\sum_{\eta=l,r}\epsilon_\eta\eta^\dagger \eta^{\phantom{\dagger}}.
\eeq
Although generic fillings can lead to a variety of interesting phases in the presence of interactions, potentially connected to fractional topological insulators~~\cite{topological_falt bands}, we shall only be concerned in this work with half-filling, i.e. $N$ fermions, where the groundstate of Eq.~\eqref{eq:FB_2q} is two-fold degenerate
\beq
\label{ni_groundstates}
\begin{split}
\ket{\epsilon_{\rm g},{\rm L}}_{\rm NI}=l^\dagger w_{1,-}^{\dagger}w_{2,-}^{\dagger}\cdots w_{N-1,-}^{\dagger}\ket{0},\\
\ket{\epsilon_{\rm g},{\rm R}}_{\rm NI}= r^\dagger w_{1,-}^{\dagger}w_{2,-}^{\dagger}\cdots w_{N-1,-}^{\dagger} \ket{0},
\end{split}
\eeq
and the groundstate energy is 
\beq
\label{eq:NI_energy}
\epsilon_{\rm g,L}=\epsilon_{\rm g,R}=-2\tilde{t}(N-1).
\eeq 
We thus see that the groundstate degeneracy corresponds to the two possible choices in populating either of the zero-energy edge modes, and it is related to the topology of the ladder (i.e. ring versus line with open edges). 

The effects of the leg imbalance $\Delta\epsilon>0$ in~\eqref{eq:creutz-hubbard_pert} can be understood from this edge perspective by writing 
\beq
\label{eq:imb}
\begin{split}
V_{\rm imb}&= \sum_{i=2}^{N-1} t_{\rm imb}\left(w_{i-1,+}^\dagger-w_{i-1,-}^\dagger \right) \left(w_{i,+}^{\phantom{\dagger}}+w_{i,-}^{\phantom{\dagger}}\right)\\
&+\sum_{\alpha=\pm}\sqrt{2}t_{\rm imb}\left(- l^{\dagger}w_{1,\alpha}^{\phantom{\dagger}}-\ii\alpha r^{\dagger}w_{N-1,\alpha}^{\phantom{\dagger}}\right)+{\rm H.c.},
\end{split}
\eeq 
where $t_{\rm imb}= -\ii\Delta\epsilon/4$ is an effective tunnelling induced by the imbalance, and has two relevant effects. The first line describes the hopping of fermions in neighboring AB-c, which leads to the aforementioned curvature of the bulk energy bands~\eqref{eq:bulk_bands}. The second line represents the hopping between the topological edge modes and the bulk AB-c. As discussed in more detail in Sec.~\ref{sec:int_regime}, both terms conspire to induce a broadening of the edge modes in the regime $4\tilde{t}\leq\Delta\epsilon$, which is the regime where the topological Zak's phase~\eqref{eq:zak_phase} vanishes, signaling an \emph{imbalance-induced topological phase transition}. We also note that, precisely at the point $4\tilde{t}=\Delta\epsilon$, the bulk bands become $\epsilon_{\pm}(q)=\pm2\tilde{t}(1+\sin q)$, and the gap vanishes exactly at $q=-\pi/2$, which coincides with the momentum of the massless Wilson fermion. 

Regarding cold atoms, previous experiments on synthetic two-leg ladders subjected to artificial gauge fluxes, where the diagonal inter-leg tunneling in Eq.~\eqref{or_cl} is substituted by a vertical one, have measured non-vanishing chiral currents circulating in opposite directions along each leg of the ladder~\cite{gauge_fields_sd}. These states are connected to the edge states of the Hofstadter model as the number of legs is increased~\cite{gauge_fields_sd,gauge_fields_sd_II}, but the bulk and edges coincide in the limiting case of the two-leg ladder. The situation is different for the two-leg Creutz ladder, as the edge states are not extended along the legs of the ladder, but exponentially localised to the left- and right-most rungs~\eqref{edge_cages}. We note that observing the particular localisation  in Eq.~\eqref{edge_cages} would require implementing a box confining potential, which have already  been achieved in several cold-atom experiments for bosonic gases~\cite{box_potentials}. However, we note that for milder confining potentials, one expects that the edge states will remain to be confined in the boundaries of the system, provided that the confining potential increases at least quadratically with the distance to the center of the trap~\cite{detection_edge_modes}.

 According to the above discussion, no additional legs would be required to identify the difference between edge and bulk excitations in the Creutz ladder, such that detecting the presence of zero-energy modes localised at the edges would be a proof of the topological nature of the phase. Using a pair of laser beams with 
a frequency difference that can be scanned within the edge-bulk gap $|\omega_{\rm L,1}-\omega_{\rm L,2}|< 2\tilde{t}$, and a spatial profile that can be localised to a few sites from the left- and right-most rungs of the ladder, the associated Bragg signal due to Raman excitations can detect the presence of these localised zero-modes, provided that the atomic gas is confined in a box potential. This method is analogous to the situation in higher-dimensional topological insulators~\cite{bragg_edge_modes}, where additional angular momentum of the laser beams can be exploited to probe the chirality of the edge modes and improve the measurement resolution~\cite{detection_edge_modes}. For milder confinement potentials, non-topological states might also happen to be localised within the edges of the system~\cite{bragg_edge_modes}. In Reference~\cite{detection_edge_modes}, the authors show that the Bragg scheme can be adapted to distinguish the chirality of the localized edge states of a higher-dimensional topological insulator, and thus differentiate them from the non-topological states, probing in this way the topological nature of the phase. It would be interesting to explore analog techniques to differentiate the edge states~\eqref{edge_cages}  of the Creutz ladder from spurious states that would arise if no box potential is implemented.

\subsection{Weak interactions: Quantum Ising ladder}
\label{sec:weak_interactions}

Let us now address the fate of this topological phase as the Hubbard repulsion is switched on. We start by exploring the weakly-interacting regime $\tilde{t}\gg V_{\rm v}$, and by noting that the two representations of the non-interacting imbalanced model $H_{\rm \pi C}=H_{\pi{\rm C}}(\tilde{t},\Delta\epsilon)$~\eqref{eq:pi_ch}, both in the original (Eqs.~\eqref{or_cl} and~\eqref{eq:H_H}) and in the plaquette (Eqs.~\eqref{eq:FB_2q} and~\eqref{eq:imb}) bases, are dual to each other. In particular, the mapping between spinors 
\beq
\Psi_i=(c_{i,\rm u},c_{i,\rm d})^{\rm t}\to W_i=\ee^{-\ii\frac{\pi}{4}\sigma_i^z}(w_{i,+},w_{i,-})^{\rm t},
\eeq 
induces a duality transformation $H_{\rm \pi C}(\tilde{t},\Delta\epsilon)\to\frac{\Delta\epsilon}{4\tilde{t}}H_{\rm \pi C}\big(\tilde{t},\frac{16\tilde{t}^2}{\Delta\epsilon}\big)$ with a self-dual point $\Delta\epsilon=(4\tilde{t})^2/\Delta\epsilon$ corresponding to the previous critical point $\Delta\epsilon=4\tilde{t}$. 

Such a 
 duality is reminiscent of the one occurring in the quantum Ising model (QIM)~\cite{duality_qim}, and suggests a possible equivalence between both models. Indeed, a formal equivalence is found by introducing the following rung operators
\beq
\label{rung_basis}
\begin{split}
 r_{j,1}&=\frac{\ii^j}{\sqrt{2}}\left(\ii c_{j,\rm u} + (-1)^j c_{j,\rm d}^\dagger\right),\\
 r_{j,2} &= \frac{\ii^j}{\sqrt{2}}\left( c_{j,\rm u} +\ii (-1)^j c_{j,\rm d}^\dagger\right).
 \end{split}
\eeq
Under this canonical transformation, the Hamiltonian is transformed onto
\begin{align}
\begin{split}
 H_{\rm \pi C}=&-\tilde{t}\sum_j\sum_{n=1,2} \left(r_{j,n}^\dagger r_{j+1,n}^{\phantom{\dagger}} + r_{j,n}^\dagger r_{j+1,n}^{{\dagger}} + \text{H.c.}\right) +\\
 &+ \frac{\Delta\varepsilon}{4}\sum_j\sum_{n=1,2} \left(2r_{j,n}^\dagger r_{j,n}^{\phantom{\dagger}} - 1\right).
\end{split}
\end{align}
In this particle-hole rung basis~\eqref{rung_basis}, we identify two independent subsystems which no longer display particle-number conservation, but instead have parity conservation. A Jordan-Wigner transformation~\cite{jw}, namely
\beq
r_{j,n}^{\dagger}=\prod_{i<j}(-\sigma^{z}_{i,n})\sigma_{j,n}^+=(r_{j,n}^{\phantom{\dagger}})^{\dagger},\hspace{2ex} r_{j,n}^{\dagger}r_{j,n}^{\phantom{\dagger}}=\half\sigma_{j,n}^z+\half,
\eeq
 reveals the Ising nature of the two subsystems, and leads to a Hamiltonian that can be understood as a \emph{two-leg quantum Ising ladder}
\begin{align}
 H_{\rm \pi C}= \sum_j\sum_{n=1,2} \left(-\tilde{t} \sigma_{j,n}^x \sigma_{j+1,n}^x + \frac{\Delta\varepsilon}{4} \sigma_{j,n}^z\right).
 \label{eq:two_copy_ising}
\end{align}
For open boundary conditions, this description allows an alternative interpretation of the edge-state behaviour by writing the Ising model as a Kitaev-Majorana chain~\cite{kitaev_model}. The ferromagnetic regime is associated with two uncoupled Majorana zero-energy modes on the opposing edges of the system for each of the legs of the Ising ladder. Combining the two free Majoranas on either edge of the Ising-ladder then yields the local fermionic edge modes defined in Eq.~\eqref{edge_cages}. This contrasts the case of a single Majorana chain, where the fermionic zero mode is highly non-local since it can only be built from the Majoranas at the opposite edges of the chain.

Let us turn on the interactions, and express the weak Hubbard repulsion in terms of these rung operators 
\beq
\begin{split}
 V_{\rm Hubb} = & \frac{V_{\rm v}}{4}\sum_j \big(\ii r_{j,1}^\dagger r_{j,2}^{\phantom{\dagger}} +{\rm H.c.}\big) \\
 - &\frac{V_{\rm v}}{4}\sum_j \big[\big(1-2r_{j,1}^\dagger r_{j,1}^{\phantom{\dagger}} \big)\big(1-2r_{j,2}^\dagger r_{j,2}^{\phantom{\dagger}}\big)-1\big].
 \end{split}
\eeq
These terms introduce a coupling between the two legs of the quantum Ising ladder. If we restrict to half-filling in the original model, the tunneling term vanishes, and we are left with the quartic interactions. These in turn can be expressed in terms of spin-spin interactions that couple the two legs of the ladder
\beq
 V_{\rm Hubb}=-\frac{V_{\rm v}}{4}\sum_j\sigma_{j,1}^z\sigma_{j,2}^z +{\rm const.}
\eeq

For weak interactions, $V_v\ll \Delta\varepsilon,\tilde{t}$, we can treat the influence of one Ising chain on the remaining one with a mean-field approximation
\begin{align}
 H_{\rm \pi CH} \approx \sum_{j,n} \left(-\tilde{t} \sigma_{j,n}^x \sigma_{j+1,n}^x + \frac{\Delta\varepsilon-V_{\rm v}m_{\bar{n}}(\Delta \varepsilon,V_{\rm v},\tilde{t}))}{4} \sigma_{j,n}^z\right),
\end{align}
where we have introduced the transverse magnetization $m_{\bar{n}}(\Delta \varepsilon,V_{\rm v},\tilde{t})=\langle \sigma_{j,\bar{n}}^z\rangle$ for each leg of the ladder, and $\bar{n}=2,1$ for $n=1,2$. We thus observe a renormalization of the imbalance parameter that controls the transverse field of the Ising model, and thus leads to a shift of the critical point as the interaction strength $V_{\rm v}$ increases. 

Accordingly, the topological phase of Sec.~\ref{sec:flat_bands} survives in a finite region of parameter space as the interactions are switched on. We find this region by solving the self-consistency mean-field equations by iterating
\begin{align}
 m_n(\Delta \varepsilon,V_v,\tilde{t}) = M\left(\frac{\Delta\varepsilon-V_v m_{\bar{n}}(\Delta \varepsilon,V_v,\tilde{t})}{4\tilde{t}}\right)
\end{align}
 to convergence. In this equation, we have used the exact expression of the ground-state transverse magnetization in the QIM~\cite{ising_magnetization}, namely
\begin{align}
 M(\alpha) &=\begin{cases}
 \frac{2(1 - \alpha^2)}{\pi \alpha} (\Pi(\alpha^2,\alpha) - K(\alpha))&\quad\text{for}\quad \alpha<1,\\
 \frac{2(\alpha^2 - 1)}{\pi \alpha^2} \Pi(1/\alpha^2,1/\alpha)&\quad\text{for}\quad \alpha>1,
 \end{cases}
\end{align}
where we have introduced the complete elliptic integrals of the first and third kind, namely
\begin{align}
 K(k) &= \int_0^\frac{\pi}{2} \dd \theta \frac{1}{\sqrt{1-k^2\sin^2 \theta}},\\
 \Pi(n,k) &= \int_0^\frac{\pi}{2} \dd \theta \frac{1}{(1+ n \sin^2 \theta) \sqrt{1-k^2 \sin^2\theta}}.
\end{align}

\begin{figure}
\centering
\includegraphics[width=0.90\columnwidth]{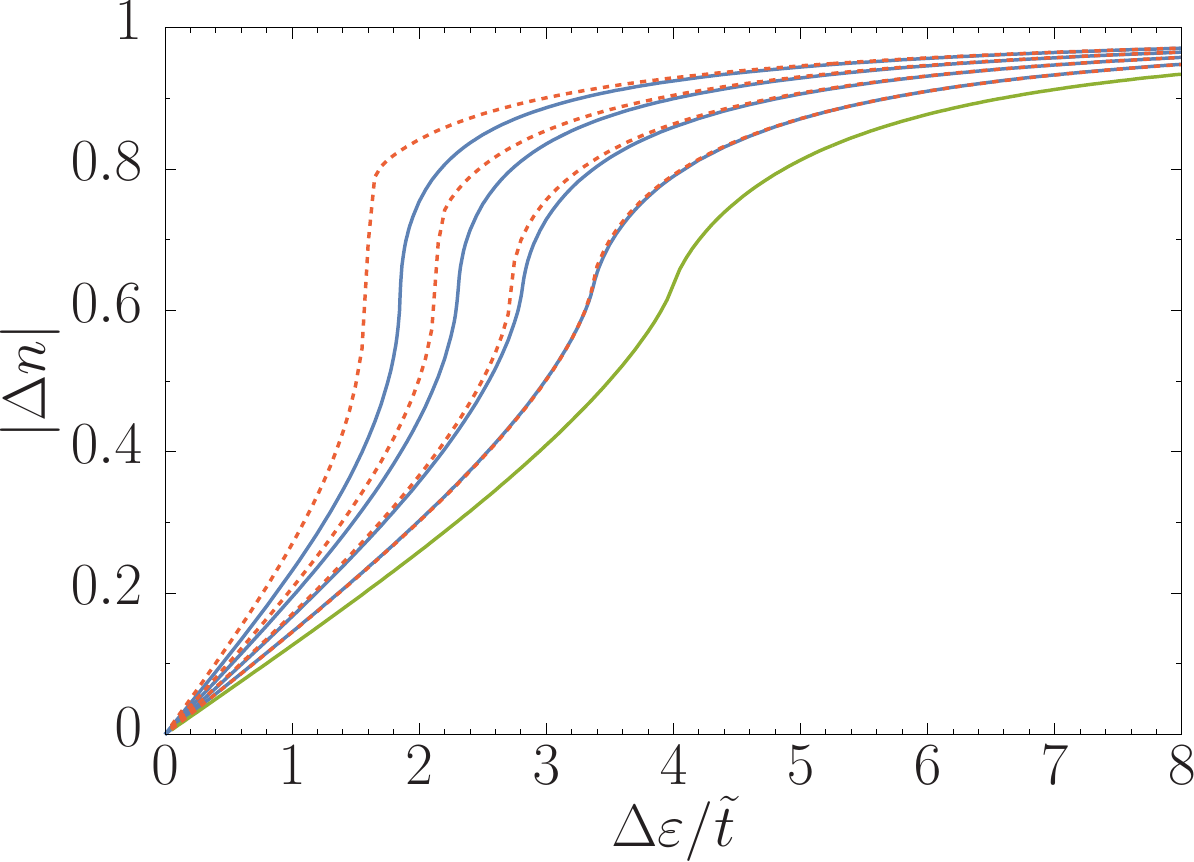}
\caption{ {\bf Exact density imbalance and mean-field magnetizations:} In blue solid lines, we present the ground-state density imbalance $\langle \Delta n\rangle$ for systems with $V_v/\tilde{t}=1,2,3,4$ (from bottom to top) as determined numerically by means of a Matrix-Product-State algorithm for a ladder with $L=256$ sites. In red dashed lines, we show the mean-field magnetisations $m_1(\Delta \varepsilon,V_v,\tilde{t})=m_2(\Delta \varepsilon,V_v,\tilde{t})$ of the coupled Ising chains. The agreement $\langle \Delta n\rangle\approx\half(m_1(\Delta \varepsilon,V_v,\tilde{t})+m_2(\Delta \varepsilon,V_v,\tilde{t}))$ is reasonable even for considerably large interaction strengths. The green curve shows as a reference the magnetization of the non-interacting model $M(\Delta\epsilon/4\tilde{t})$.}
\label{fig_mean_field}
\end{figure}

The self-consistent mean-field approximation reproduces very precisely the numerical results for the density imbalance of the Creutz ladder 
\beq
\label{eq:leg_imbalance}
 \Delta n=\frac{1}{N} \sum_i\left(c_{i,\rm u}^\dagger c_{i,\rm u}^{\phantom{\dagger}} - c_{i,\rm d}^\dagger c_{i,\rm d}^{\phantom{\dagger}}\right) = \frac{1}{2N} \sum_j\sum_n\sigma_{j,n}^z.
 \eeq
 In Fig.~\ref{fig_mean_field}, we compare the groundstate density imbalance $\langle \Delta n\rangle$ with the mean-field transverse magnetization $m_{1}(\Delta \varepsilon,V_{\rm v},\tilde{t})=m_{2}(\Delta \varepsilon,V_{\rm v},\tilde{t})$, and find a very good agreement even for considerable interactions $V_{\rm v}\sim \tilde{t}$.
 
If we solve the self-consistency equation to first order in $V_v/\tilde{t}$, we find that the critical point $ \Delta\epsilon/\tilde{t} = 4 $ flows towards smaller values of the imbalance as the interactions increase
 \begin{align}
 \label{eq:critical_line_mean_field}
 \frac{\Delta\epsilon}{\tilde{t}} = 4 -\frac{2}{\pi} \frac{V_v}{\tilde{t}} + \mathcal{O}\left(\frac{V_v^2}{\tilde{t}^2}\right).
\end{align}
This weak-coupling expansion defines a critical line in parameter space that separates the topological and non-topological phases, and agrees well with our numerical findings for the phase diagram of the model, as discussed below (see the red dashed line in  Fig.~\ref{fig_phase_diagram}).

Regarding the cold-atom experiment, we note that the definition of the Zak phase~\eqref{eq:zak_phase}, which could be used to probe the topological nature of the phase via interferometric methods~\cite{zak_phase_other_models}, is only valid in the non-interacting limit. For weak interactions, the single-particle excitations will be described instead by quasi-particles, and the interferometric protocol to measure a many-body Zak's phase can be obtained by coupling additional impurities to the interacting Fermi gas~\cite{many_body_zak_phase}. We also note that the method to detect the zero-energy edge modes based on Bragg scattering~\cite{bragg_edge_modes}, should also hold in the interacting regime. Finally, we advance that a measurement of the leg imbalance density~\eqref{eq:leg_imbalance} could also be used to experimentally test the validity of our prediction of the critical line~\eqref{eq:critical_line_mean_field} (see Sec.~\ref{sec:MPS}). Concerning the mapping in Eq.~\eqref{eq:synth_dim}, this leg imbalance can be inferred from spin-resolved density 
measurements obtained by optical imaging, either in-situ (dispersive imaging) or after a time-of-flight expansion (absorption imaging).  Exploiting light polarization and selection rules for the particular internal states of the atoms, one could measure separately the density of each of the two components, and even spin structure factors~\cite{bragg_spin_structure} that give access to the  leg-imbalance susceptibility. Note that this susceptibility can be used to extract the position of the critical line (see top panel of  Fig.~\ref{fig:ti-opm}). Regarding in-situ absorption imaging, spin-selective density measurements and correlations can also be accomplished~\cite{absorption_correlations}. As  alternative  possibilities, the spin structure factor could be measured in a non-demolition manner  by exploiting  a quantum Faraday effect~\cite{q_faraday}, while spin-resolved  density measurements can be achieved by using  recent quantum gas microscopes (see e.g.~\cite{spin_microscope}).

\subsection{Strong interactions: Orbital Ising ferromagnet}
\label{sec:XYmodel}

In this section, we explore the opposite limit of the Creutz-Hubbard ladder~\eqref{eq:pi_ch}, namely the strongly-interacting regime $\tilde{t}\ll V_{\rm v}$. In the limit $\tilde{t}=0$, the groundstate of the Creutz-Hubbard Hamiltonian corresponds to a Mott insulator where the $N$ fermions are distributed in the ladder avoiding simultaneous occupancies of two sites within the same rung. 

By switching on the tunnelling $\tilde{t}\ll V_{\rm v}$, an orbital analog of the well-known super-exchange~\cite{superexchange_anderson,superexchange_theory,superexchange_bhm_exp,superexchange_fhm} takes place, which reduces the energy of the groundstate by allowing for processes where a rung of the ladder is virtually occupied by two fermions. Instead of the usual Heisenberg model associated to the strongly-interacting Hubbard model~\cite{superexchange_anderson}, a different spin model arises in the present case due to the combination of intra- and inter-leg tunnellings. 
To second order of perturbation theory~\cite{t_U_expansion}, and considering the half-filling regime, the relevant Hamiltonian describing the low-energy physics corresponds to an \emph{orbital quantum Ising model}, namely
\beq
\label{eq:Ising}
\mathcal{P}_{\rm r}H_{\rm \pi CH}\mathcal{P}_{\rm r}=\fourth JN+J\sum_{i}T_i^yT_{i+1}^y+\Delta\epsilon \sum_iT_i^z,
\eeq
where the super-exchange coupling is $J=-8\tilde{t}^2/V_{\rm v}$, and the leg imbalance $\Delta\epsilon$ plays the role of an effective transverse field. The above spin operators are defined as the orbital analogue of the usual spin operators for electrons
\beq
\label{si_spins}
T_i^y= \frac{1}{2} \left(-\ii c_{i,\rm u}^\dagger c_{i,\rm d}^{\phantom{\dagger}}+\ii c_{i,\rm d}^\dagger c_{i,\rm u}^{\phantom{\dagger}}\right),\hspace{1ex} T_i^z=\frac{1}{2} \left(c_{i,\rm u}^\dagger c_{i,\rm u}^{\phantom{\dagger}}-c_{i,\rm d}^\dagger c_{i,\rm d}^{\phantom{\dagger}}\right).
\eeq
Finally, $\mathcal{P}_{\rm r}=\Pi_i(1-n_{i,\rm u}n_{i,\rm d})$ is a Gutzwiller projector onto the subspace of singly-occupied vertical rungs.

The 1D quantum Ising model can be exactly solved~\cite{Ising_model} by introducing a Jordan-Wigner transformation~\cite{jw},
followed by a fermionic Bogoliubov transformation~\cite{bogoliubov}. In comparison with the non-interacting groundstate~\eqref{ni_groundstates}, which displays a topological two-fold degeneracy, the strongly-interacting groundstate for $\Delta\epsilon<|J|/2$ has a non-topological degeneracy related to the $\mathbb{Z}_2$ symmetry of the Ising model. In this regime, the groundstate develops long-range order as a consequence of spontaneous symmetry breaking 
\beq
\label{eq:ferro_orbital_order}
{\rm lim}_{r\to\infty }\langle T_{i}^yT_{i+r}^y\rangle_{\rm SI}= \fourth\left(1-h^2\right)^{\fourth},
\eeq 
where $h=2\Delta\epsilon/|J|<1$. This defines a critical line
\beq
\label{eq:phase_boundary_Ising}
\frac{\Delta\epsilon}{\tilde{t}}=\frac{4\tilde{t}}{V_{\rm v}},
\eeq
that separates the phase of long-range order, i.e. an \emph{orbital ferromagnet} (oFM), from the disordered phase, i.e. an \emph{orbital paramagnet} (oPM), and is depicted by a yellow dashed line in  Fig.~\ref{fig_phase_diagram}. As the leg imbalance is increased above a critical value $\frac{\Delta\epsilon}{\tilde{t}}|_{\rm c}=\frac{4\tilde{t}}{V_{\rm v}}$, a standard quantum phase transition occurs between the long-range ordered Ising ferromagnet and a disordered orbital paramagnet, where all fermions tend to occupy the lower leg of the ladder. We note that this transition is not of a topological origin, as it can be understood by a local order parameter: the orbital magnetisation $\langle T_{i}^yT_{i+r}^y\rangle_{\rm SI}\to m_y^2$.

We also note that the long-range ferromagnetic order is totally absent in the non-interacting topological ground-state~\eqref{ni_groundstates}, where one finds
\beq
\langle T_{i}^yT_{i+r}^y\rangle_{\rm NI}=0.
\eeq 
According to this discussion, it is clear that one cannot connect the non-interacting topological and strongly-interacting ferromagnetic phases adiabatically. Therefore, there should be an \emph{interaction-induced topological quantum phase transition} between the symmetry-protected topological phase, and a state with magneto-orbital long-range order, for intermediate interactions $V_{\rm v}/\tilde{t}$. Our analytical treatment of the strongly-interacting regime also points to the possible origin of in-plane ferromagnetic order in the Creutz-Hubbard ladder with vertical tunnelings~\eqref{or_cl} instead of the imbalance, as recently found through a mean-field and numerical analysis~\cite{creutz_majorana_hubbard}. Similar topological quantum phase transitions to magnetically-ordered phases have also been found numerically in higher-dimensional models~\cite{interacting_ti_review}.

Concerning the cold-atom realization, we note that the orbital spin operators~\eqref{si_spins} developing the long-range order~\eqref{eq:ferro_orbital_order} in the orbital ferromagnet, correspond to the standard spin operators for the two internal states of the atoms~\eqref{eq:synth_dim}. Therefore, the magnetic correlations can be inferred from the spin structure factor, which can be measured for instance via Bragg scattering by playing with the polarization of the laser beams~\cite{bragg_spin_structure}. This can allow to test experimentally the validity of the critical line~\eqref{eq:phase_boundary_Ising}.

\subsection{Intermediate interactions: Extended Hubbard models and quantum impurity physics}
\label{sec:int_regime}

\begin{figure*}
\centering
\includegraphics[width=1.9\columnwidth]{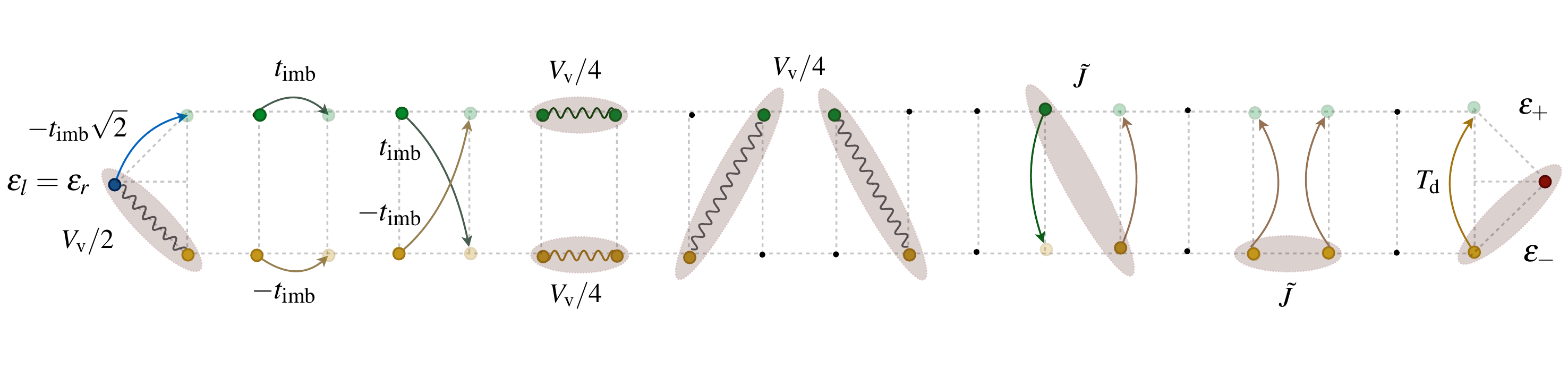}
\caption{ {\bf Non-standard Hubbard terms in the Aharonov-Bohm-cage basis: } Virtual ladder structure to represent the relevant processes in the Creutz-Hubbard ladder. The two legs of the ladder correspond to the two possible flat bands at energies $\epsilon_\pm=\pm2\tilde{t}$, with sites representing the bulk Aharonov-Bohm cages~\eqref{ab_cages}. The two boundary sites represent the two zero-energy modes $\epsilon_{l}=\epsilon_r=0$, and their corresponding edge cages~\eqref{edge_cages}.
 The imbalance of the original Creutz-Hubbard ladder~\eqref{eq:imb} leads to standard intra- and inter-leg tunnelling, but also to edge-bulk tunnelling processes represented by curved arrows and with strengths proportional to $t_{\rm imb}$. The Hubbard interaction~\eqref{ne_FH} can be decomposed into nearest-neighbour interactions of strength proportional to $V_{\rm v}$ that are represented by curly lines, and involve two adjacent bulk cages in any possible leg, or adjacent edge-bulk ones. Non-standard Hubbard terms involving a pair tunnelling also arise, where pairs of fermions tunnel simultaneously either from the same leg (correlated tunnelling) or from opposite legs (anti-correlated tunnelling) of the ladder with amplitude $\tilde{J}$. Finally, we also describe a density-dependent inter-leg tunnelling of amplitude amplitude $T_{\rm d}$, where the fermion tunnelling depends on the density of fermions populating the Aharonov-Bohm cage at the neighbouring edge. }

\label{fig_ne_FH_scheme}
\end{figure*}

In this section, we elaborate on a theory that allows us to predict this interaction-induced topological quantum phase transition by making an interesting connection to the physics of quantum impurity models.
We start by expressing the Hubbard part of the Hamiltonian~\eqref{eq:creutz-hubbard_pert} in the basis of the bulk~\eqref{ab_cages} and edge~\eqref{edge_cages} Aharonov-Bohm cages (AB-c). A similar step has been performed by rewriting the bosonic Creutz-Hubbard model in terms of the bulk negative-energy AB-c~\cite{bosonic_creutz_hubbard_aoki,bosonic_creutz_hubbard_huber}. This corresponds to a projection onto the low-energy flat band, which is justified for $\tilde{t}\gg V_{\rm v}$, and is also customary in the theory of fractional topological insulators~\cite{topological_falt bands}. In our case, however, it will be crucial to retain the positive- and zero-energy AB-c to capture the topological phases, and the possible quantum phase transitions that take place in the regime $\tilde{t}\sim V_{\rm v}$. 

In this basis, we find an extended Hubbard model
\beq
\label{ne_FH}
V_{\rm Hubb}=V_{\rm nn}+V_{\rm pt}+V_{\rm dt},
\eeq
with a \emph{nearest-neighbour interaction} between fermions confined in adjacent AB-c
\beq
\label{eq:nn}
V_{\rm nn}=\frac{V_{\rm v}}{2}\!\sum_{\alpha=\pm}\!\!\left(n_l n_{1,\alpha}+n_rn_{N-1,\alpha}\right)+\frac{V_{\rm v}}{4}\sum_{i=2}^{N-1}\!\sum_{\alpha,\beta=\pm}\!n_{i-1,\alpha} n_{i,\beta},
\eeq
where the AB-c number operators are $n_{i,\alpha}=w^{\dagger}_{i,\alpha}w^{\phantom{\dagger}}_{i,\alpha}$ for $\alpha\in\{+,-\}$, and $n_\eta=\eta^\dagger\eta^{\phantom{\dagger}}$ for $\eta\in\{\ell, r\}$. This effective repulsion between the fermions can be depicted in a new ladder scheme, where the legs of the ladder correspond to the positive/negative energy flat bands, the sites correspond to the labels of the different bulk cages, except at the boundaries of the ladder, where the sites correspond to the edge cages (see Fig.~\ref{fig_ne_FH_scheme}). All the possible interactions, including the ones at the edges, are represented by curly lines. We also include the nearest-neighbour intra- and inter-leg tunnelling, which arises from interpreting the imbalance~\eqref{eq:imb} within this virtual ladder.

In addition to this effective tunnelling and repulsion, we also find \emph{correlated tunnelling} processes of two types. We obtain 
a \emph{pair-tunnelling} Hamiltonian
\beq
\label{eq:pt}
\begin{split}
V_{\rm pt}&=\tilde{J}\sum_{i=2}^{N-1} \left(w_{i-1,+}^\dagger w_{i-1,-}^{\phantom{\dagger}} \!\big(w_{i,-}^{{\dagger}}w_{i,+}^{\phantom{\dagger}}+ w_{i,+}^{{\dagger}}w_{i,-}^{\phantom{\dagger}}\big)\right)+{\rm H.c.}\\
\end{split}
\eeq
where $\tilde{J}=-V_{\rm v}/4$. The first term in the parenthesis describes how fermions confined in adjacent AB-c of opposite legs tunnel simultaneously to empty neighbouring cages, and can be understood as the anti-correlated pair-tunnelling of Fig.~\ref{fig_ne_FH_scheme}. The second term describes how fermions confined in adjacent Aharonov-Bohm cages on the same leg tunnel simultaneously to empty neighbouring cages, and can be understood as the correlated pair-tunnelling of Fig.~\ref{fig_ne_FH_scheme}. In addition, we also obtain an inter-leg \emph{density-dependent tunnelling} 
\beq
\label{eq:ddt}
\begin{split}
V_{\rm dt}&=T_{\rm d}\sum_{i=2}^{N-1}\left(n_{i-1,+}+n_{i-1,-}-n_{i+1,+}-n_{i+1,-}\right)w_{i,+}^\dagger w_{i,-}^{\phantom{\dagger}}\\
&+2T_{\rm d}\left(n_l w_{1,+}^{{\dagger}}w_{1,-}^{\phantom{\dagger}}-n_r w_{N-1,+}^{{\dagger}}w_{N-1,-}^{\phantom{\dagger}}\right) +{\rm H.c.}, 
\end{split}
\eeq
where $T_{\rm d}=V_{\rm v}/4$ is the tunneling strength. The first line describes an inter-leg tunnelling within the bulk of the ladder that depends on the density difference of the neighbouring Aharonov-Bohm cages, and will be negligible for a groundstate with translationally-invariant bulk properties. On the other hand, the second line describes inter-leg tunnelings that occur at the boundaries of the Creutz-Hubbard ladder, and depend on the density of the edge AB-c (see Fig.~\ref{fig_ne_FH_scheme}). These terms are not negligible for translational-invariant bulks, and will play a key role in the topological phase transitions of the model.

As shown in this section, the flat-band physics of the Creutz-Hubbard ladder provides an alternative route to the physics of non-standard Hubbard models, which typically arise in optical lattices as a consequence of dipolar interactions or higher orbitals~\cite{non_standard_hubbard}. In the present case, these non-standard terms arise due to the interplay of interactions and the flatness of the bands, which yield
\beq
\label{eq:ch_ab_basis}
H_{\rm \pi CH}=H_{\rm FB}+V_{\rm imb}+V_{\rm nn}+V_{\rm pt}+V_{\rm dt},
\eeq
with the terms introduced in Eqs.~\eqref{eq:FB_2q},~\eqref{eq:imb}, and~\eqref{eq:nn}-\eqref{eq:ddt}. At first sight, this formulation leads to a Hamiltonian~\eqref{eq:ch_ab_basis} that seems more complicated than the original one~\eqref{eq:pi_ch}. However, as shown below, it is an ideal starting point to calculate an effective boundary theory for the topological edge states, based on which one can understand possible topological phase transitions as the imbalance and/or interactions are increased.

\subsubsection{Bulk-mediated broadening and edge-to-edge tunnelling}

As a warm-up to the discussion of the model for intermediate interactions, let us discuss how to derive an effective boundary theory in the non-interacting limit $V_{\rm v}=0$. In Sec.~\ref{sec:flat_bands}, we have already discussed how the imbalance~\eqref{eq:imb} may induce a quantum phase transition at $\frac{\Delta\epsilon}{\tilde{t}}|_{\rm c}=4$ that can be understood through a topological invariant~\eqref{eq:zak_phase} of the bulk bands. We now discuss how an effective boundary theory, derived via the above non-standard Hubbard Hamiltonian~\eqref{eq:ch_ab_basis}, yields an alternative description of such a phase transition.

 The Hamiltonian~\eqref{eq:pi_ch} can be rearranged as follows $H_{\pi\rm C}=H_{\rm edge}+H_{\rm bulk}+H_{\rm b-e}$, where
\beq
\label{edge}
H_{\rm edge}=\epsilon_l l^\dagger l^{\phantom{\dagger}}+\epsilon_r r^\dagger r^{\phantom{\dagger}},
\eeq
stands for the Hamiltonian of the zero-energy edge modes. The bulk part of the Hamiltonian can be expressed as follows
\beq
\label{bulk_lattice}
H_{\rm bulk}=\sum_{i,\alpha}\!\epsilon_\alpha w_{i,\alpha}^\dagger w_{i,\alpha}^{\phantom{\dagger}}+\sum_{i=2}^{N-1}\!\! t_{\rm imb}(w_{i-1,+}^\dagger-w_{i-1,-}^\dagger ) (w_{i,+}^{\phantom{\dagger}}+w_{i,-}^{\phantom{\dagger}}),
\eeq
which is readily diagonalised by a Fourier transform considering periodic boundary conditions $ w_{N,\alpha}=w_{1,\alpha}$. This leads to the energy bands in Eq.~\eqref{eq:bulk_bands}, where the quasi-momentum lies in the Brillouin zone $q\in(-\pi,\pi]$, and there is a total of $2N$ single-particle modes. To approximately solve the two-leg Creutz ladder with open boundary conditions, we build on the solution of an open single chain. We note that the bands in our ladder fulfill $\epsilon_\alpha(\frac{\pi}{2}+q)=\epsilon_\alpha(\frac{\pi}{2}-q)$, which allows us to combine modes with momenta $\frac{\pi}{2}\pm q$ to construct AB-c momentum operators via standing waves that respect the open boundary conditions of a finite ladder $ w_{N,\alpha}=w_{0,\alpha}=0$.
Note that the $-q$ solutions of the problem with periodic boundary conditions are implicitly considered in the $+q$ solutions. Together with the fact that the $q=0,\pi$ modes of the periodic solution yield trivial standing waves, this forces us to enlarge the number of sites in the problem with periodic boundary conditions $N-1\to 2N$, such that the allowed momenta become $q_n=\frac{2\pi}{N-1}n\to q_n=\frac{\pi }{N}n$. The transformation that approximately diagonalises the bulk Hamiltonian~\eqref{bulk_lattice} is
\beq
\label{ft}
\left(\!\begin{array}{c}w_{+}(q_n) \\w_{-}(q_n)\end{array}\!\right)\!=\sqrt{\frac{2}{N}}\sum_{j=1}^{N-1}\!\ee^{-\ii \frac{\pi}{2}j}\sin (q_n j)\left(\!\begin{array}{c} w_{j,+} \\ w_{j,-}\end{array}\!\right).
 \eeq
Using these operators, the bulk Hamiltonian for the ladder with open ends  approximately becomes
\beq
H_{\rm bulk}\approx\sum_{n=1}^{N-1}{\Psi}^\dagger(q_n)\boldsymbol{B}(q_n)\cdot\boldsymbol{\sigma}\hspace{0.2ex}{\Psi}(q_n),
\eeq
where ${\boldsymbol{B}}(q_n)=(0,\frac{\Delta\epsilon}{2}\cos q_n,2\tilde{t}+\frac{\Delta\epsilon}{2}\sin q_n)$, and we have introduced new spinors ${\Psi}(q_n)=(w_{+}(q_n),w_{-}(q_n))^{\rm t}$. Diagonalising this Hamiltonian leads to the curved energy bands~\eqref{eq:bulk_bands}, such that $ H_{\rm bulk}=\sum_{n,\alpha}\alpha\epsilon(q_n)\gamma^{\dagger}_\alpha(q_n)\gamma^{\phantom{\dagger}}_\alpha(q_n)$~\cite{comment_edge_bulk}, where we have introduced
\beq
\label{eq:imbalance_diag_operators}
\begin{split}
\gamma^{\phantom{\dagger}}_+(q_n)&=+u_{q_n}^*w_+(q_n)+v_{q_n} w_-(q_n),\\
\gamma^{\phantom{\dagger}}_-(q_n)&=-v_{q_n}^*w_+(q_n)+u_{q_n} w_-(q_n),\\
\end{split}
\eeq
together with the following constants
\beq
u_{q_n}=\sqrt{\frac{\epsilon(q_n)+\zeta(q_n)}{2\epsilon(q_n)}},\hspace{2ex} 
v_{q_n}=\ee^{\ii\phi_n} \sqrt{\frac{\epsilon(q_n)-\zeta(q_n)}{2\epsilon(q_n)}},
\eeq
and $\zeta(q_n)=2\tilde{t}+\frac{\Delta\epsilon}{2}\sin q_n$, $\ee^{\ii\phi_n}=-\ii \, {\rm sgn}(\cos(q_n))$. We note that by using the periodic  bulk energies~\eqref{eq:bulk_bands}, our approximation is essentially introducing an intensive contribution coming from the added bonds at the ends of the ladder.

Finally, the imbalance~\eqref{eq:imb} also induces a bulk-edge coupling that can be understood as a hybridization between the edge and bulk orbitals 
\beq
\label{eq:hybridization}
H_{\rm b-e}=\sum_{n,\alpha} (g^l_{n,\alpha}l^\dagger +g^r_{n,\alpha}r^\dagger )\gamma_\alpha(q_n)+{\rm H.c.}.
\eeq
To obtain the correct bulk-edge couplings $g^{\eta}_{n,\alpha}$, note that the connection between the solutions of the periodic and open chains required enlarging the number of sites of the periodic chain. To preserve the distance between the edges, one must  change the site indexes of operators that form the edge-bulk coupling~\eqref{eq:imb}, and leads to 
 \beq
 \begin{split}
 g_{n,+}^l&=-\frac{\Delta\epsilon}{2\sqrt{N}}\sin q_n(u_{q_n}-v_{q_n}), \\
 g_{n,-}^l&=-\frac{\Delta\epsilon}{2\sqrt{N}}\sin q_n(u^*_{q_n}+v^*_{q_n}), \\
 g_{n,+}^r&=\frac{\Delta\epsilon}{2\sqrt{N}}(-\ii)^{\frac{N-1}{2}}\sin \left(\half q_n(N-1)\right)(u_{q_n}+v_{q_n}),\\
 g_{n,-}^r&=\frac{\Delta\epsilon}{2\sqrt{N}}(-\ii)^{\frac{N-1}{2}}\sin \left(\half q_n(N-1)\right)(v^*_{q_n}-u^*_{q_n}).\\
 \end{split}
 \eeq 
 Interestingly, the Hamiltonian of the Creutz ladder in this formulation $H_{\rm edge}+H_{\rm bulk}+H_{\rm b-e}$ is similar to a non-interacting two-impurity \emph{Fano-Anderson model}~\cite{anderson_fano_model}, where the impurities correspond to the modes localised at the edges of the Creutz ladder, and the gapless metallic bands are substituted by gapped bulk bands also described by free fermions. This analogy yields an insightful interpretation of the topological quantum phase transition at $\frac{\Delta\epsilon}{\tilde{t}}|_{\rm c}=4$.

Interpreting the dispersive bulk bands as reservoirs, the hybridisation can have two effects: \emph{(i)} the zero-energy edge modes $\eta=l,r$ may get shifted and broadened $\epsilon_\eta=0\to\Delta\epsilon_\eta-\ii\Gamma_\eta/2$, \emph{(ii)} the bulk fermions mediate an edge-to-edge tunnelling. Due to the particle-hole symmetry of the bands, the level shifts cancel $\Delta\epsilon_\eta=0$, such that the poles representing the edge excitations always correspond to zero energy. On the other hand, the broadening can be expressed as $\Gamma_\eta=\sum_\alpha J_{\eta,\alpha}(\epsilon_\eta)$, where we have introduced the spectral density for the coupling of the edge states to the bulk bands
\beq
\label{eq:hybridization_spectral_density}
J_{\eta\alpha}(\omega)=2\pi\sum_n|g_{n,\alpha}^\eta|^2\delta(\omega-\alpha\epsilon(q_n)).
\eeq
Accordingly, the broadening of the levels depends on the value of the spectral function at zero energy $\Gamma_\eta=\sum_\alpha J_{\eta\alpha}(0)$. For $\Delta\epsilon<4\tilde{t}$, the bulk bands~\eqref{eq:bulk_bands} remain gapped, and thus $J_{\eta\alpha}(0)=0$, such that $\Gamma_l=\Gamma_r=0$, and the edge modes thus remain in-gap bound states. As announced before, there will be a bulk-mediated tunnelling 
between these in-gap modes
 \beq
 \label{eq:ee_tunneling}
 H_{\rm e-e}=t_{\rm ee}r^\dagger l+{\rm H.c.},
 \eeq
 where we have introduced the tunnelling strength
 \beq
 t_{\rm ee}= \sum_{n,\alpha}\frac{g_{n,\alpha}^r\left(g_{n,\alpha}^l\right)^*}{ \alpha\epsilon(q_n)}.
 \eeq
 Provided that the bulk bands~\eqref{eq:bulk_bands} remain gapped, this edge-edge tunnelling will decrease exponentially with the edge-to-edge distance, such that the topological degeneracy of the groundstates~\eqref{ni_groundstates} is preserved in the thermodynamic limit. Conversely, when $\Delta\epsilon =4\tilde{t}$, the gap vanishes and the exponential localisation of the edge states disappears. Moreover, the bulk density of states at zero energy does not vanish anymore, such that $J_{\eta\alpha}(0)\neq0$, and the edge modes become broadened resonances $\Gamma_\eta>0$ instead of the previous bound states.

 It is very instructive to understand how the onset of a topological phase transition can be predicted by looking at the effective theory for the edges. This will become very useful in the presence of interactions, where a simple interpretation in terms of observables associated to a topological invariant~\eqref{eq:zak_phase} for non-interacting systems cannot be applied. Additionally, this result also underlies the importance of preserving all the orbitals in the effective description in terms of AB-c, and not simply projecting to the low-energy flat band when trying to describe flat-band effects in fermionic topological insulators.

\subsubsection{Bulk-mediated dephasing and edge-edge interactions}
 
Let us now derive an effective boundary theory in the balanced interacting model ($\Delta \varepsilon = 0$) for intermediate interactions $V_{\rm v}/\tilde{t}$. Therefore, we must analyse the following part $H_{\rm \pi CH}=H_{\rm FB}+V_{\rm nn}+V_{\rm pt}+V_{\rm dt}$ of the Hamiltonian~\eqref{eq:ch_ab_basis}.
 Although the pair and density-dependent tunnellings in Eqs.~\eqref{eq:pt}-\eqref{eq:ddt} modify the distribution of particle-hole pairs in the rungs of the virtual ladder (see Fig.~\ref{fig_ne_FH_scheme}), the nearest-neighbour interactions~\eqref{eq:nn} do not change, since the number of neighbouring AB-c that are occupied is preserved under such processes. Therefore, in this limit, the nearest-neighbour interactions~\eqref{eq:nn} can be substituted by a $c$-number, and only the flat-band~\eqref{eq:FB_2q} and the correlated-tunnelling terms~\eqref{eq:pt}-\eqref{eq:ddt} have an important effect on the non-interacting groundstates~\eqref{ni_groundstates}. Moreover, as argued below Eq.~\eqref{eq:ddt}, only the density-dependent tunnelling at the edges of the ladder will play a role to determine the order of the translationally-invariant groundstates.
 
 According to this discussion, we can rearrange these terms as $H_{\rm \pi CH}=H_{\rm edge}+H_{\rm bulk}+H_{\rm b-e}$, where the edge Hamiltonian coincides with Eq.~\eqref{edge} for the non-interacting imbalanced case. In contrast, the bulk Hamiltonian is no longer described by a quadratic fermionic model, but instead by quartic terms that can be understood as some effective spin exchange. This becomes apparent after introducing the $\mathfrak{su}(2)$ algebra
 \beq
\tilde{T}_i^x= \frac{1}{2}\! \left(\!w_{i,\rm +}^\dagger w_{i,\rm -}^{\phantom{\dagger}}+ w_{i,\rm -}^\dagger w_{i,\rm +}^{\phantom{\dagger}}\!\right),\hspace{1ex} \tilde{T}_i^z=\frac{1}{2} \!\left(\!w_{i,\rm +}^\dagger w_{i,\rm +}-w_{i,\rm -}^\dagger w_{i,\rm -}^{\phantom{\dagger}}\!\right),
\eeq
which should not be confused with the strong-coupling orbital spin operators of Eq.~\eqref{si_spins}. With this notation, the bulk Hamiltonian becomes
 \beq
 \label{eq:eff_ising}
 H_{\rm bulk}=\frac{V_{\rm v}}{4}N+\sum_{i=1}^{N-1}4\tilde{t}\tilde{T}_i^z+\sum_{i=2}^{N-1}4\tilde{J}\tilde{T}_{i-1}^x\tilde{T}_i^x,
 \eeq
 where $\tilde{J}=-V_{\rm v}/4$, which corresponds to a ferromagnetic Ising model in a transverse field. This model can be solved exactly for periodic boundary conditions~\cite{Ising_model} by means of a Jordan-Wigner transformation~\cite{jw}, namely
 \beq
 \tilde{T}_i^z=f_i^\dagger f_i^{\phantom{\dagger}}-\half, \hspace{2ex} \tilde{T}_i^x=\half f_i^\dagger \ee^{\ii\pi\sum_{j<i}f_j^\dagger f_j^{\phantom{\dagger}}}+{\rm H.c.},
 \eeq
 where $f_i^{\dagger}=(f_i)^\dagger$ are spinless fermionic operators. Considering periodic boundary conditions $ f_{N}=f_{1}$, we would obtain the energy bands 
for single-particle excitations 
 \beq
 \label{eq:bulk_bands_ising}
 \tilde{\epsilon}_\pm(q)=\pm \tilde{\epsilon}(q)=\pm2|\tilde{J}|\sqrt{1+\tilde{\frak{f}}^{2}-2\tilde{\frak{f}}\cos q},
 \eeq
 where we have introduced the flatness parameter $\tilde{\frak{f}}=8\tilde{t}/V_{\rm v}$, such that we recover perfect flat bands in the non-interacting regime $V_{\rm v}=0$. Here, the quasi-momentum is defined within a halved Brillouin zone $q\in(0,\pi]$, such that there is a total of $N$ single-particle modes. Let us remark that, just by looking at this effective theory of the bulk, one would predict a standard quantum phase transition at $|J|=2\tilde{t}$ onto an orbital ferromagnet that can be described by a local order parameter. However, this theory would predict a featureless disordered phase for $|J|<2\tilde{t}$, and thus completely miss the topological features of the model. It is thus of paramount importance to include the boundary AB-c in the full description.
 
 To prepare for that, we use a similar approximation to treat the bulk as in the previous section. We note that the above energies have the symmetry 
 $\tilde{\epsilon}_\alpha(+q)=\tilde{\epsilon}_\alpha(-q)$, such that we can move back to the entire Brillouin zone, and construct AB-c momentum operators that respect the open boundary conditions $ f_{N}=f_{0}=0$ of a finite ladder by combining the $\pm q$ solutions of the periodic chain. In this case, in order to recover the $N$ single-particle modes, and taking into account that the $q=\pi$ mode yields a trivial sanding wave, we need to enlarge the number of sites in the problem with periodic boundary conditions $N-1\to N$, such that the allowed momenta become $q_n=\frac{2\pi}{N-1}n\to q_n=\frac{2\pi}{N}n$ with $n\in\{1,\cdots,\half N\}$. Accordingly, the mapping
\beq
\label{ft_ising}
\left(\!\begin{array}{c}f(+q_n) \\f^{\dagger}(-q_n)\end{array}\!\right)\!=\sqrt{\frac{2}{N}}\sum_{j=1}^{N-1}\sin (q_n j)\left(\!\begin{array}{c} f_{j} \\ f_{j}^{\dagger}\end{array}\!\right),
 \eeq
transforms the bulk Hamiltonian~\eqref{eq:eff_ising} approximately into 
\beq
H_{\rm bulk}\approx\sum_{n=1}^{N/2}\tilde{\Psi}^\dagger(q_n)\boldsymbol{\tilde{B}}(q_n)\cdot\boldsymbol{\sigma}\hspace{0.2ex}\tilde{\Psi}(q_n),
\eeq
where $\tilde{\boldsymbol{B}}(q_n)=(0,-2\tilde{J}\sin q_n,4\tilde{t}+2\tilde{J}\cos q_n)$, and $\tilde{\Psi}(q_n)=(f(q_n),f^\dagger(-q_n))^{\rm t}$ are the Nambu spinors. Diagonalising this Hamiltonian leads to the bulk curved energy bands~\eqref{eq:bulk_bands_ising}, such that $ H_{\rm bulk}=\sum_{n,\alpha}\alpha\tilde{\epsilon}(q_n)\tilde{\gamma}^{\dagger}_\alpha(q_n)\tilde{\gamma}^{\phantom{\dagger}}_\alpha(q_n)$, where we have introduced
\beq
\label{eq:ising_diag_operators}
\begin{split}
\tilde{\gamma}^{\phantom{\dagger}}_+(q_n)&=+\tilde{u}_{q_n}^*f(q_n)^{\phantom{\dagger}}+\tilde{v}_{q_n} f^\dagger(-q_n),\\
\tilde{\gamma}^{\phantom{\dagger}}_-(q_n)&=-\tilde{v}_{q_n}^*f(q_n)^{\phantom{\dagger}}+\tilde{u}_{q_n} f^\dagger(-q_n),\\
\end{split}
\eeq
together with the following constants
\beq
u_{q_n}=\sqrt{\frac{\tilde{\epsilon}(q_n)+\tilde{\zeta}(q_n)}{2\tilde{\epsilon}(q_n)}},\hspace{2ex} 
v_{q_n}=\ee^{\ii\tilde{\phi}_n} \sqrt{\frac{\tilde{\epsilon}(q_n)-\tilde{\zeta}(q_n)}{2\tilde{\epsilon}(q_n)}},
\eeq
and $\tilde{\zeta}(q_n)=2\tilde{t}+\tilde{J}\cos q_n$, $\ee^{\ii\tilde{\phi}_n}=\ii {\rm sgn}(\sin(q_n))$. In contrast to the single-particle bulk excitations in the imbalanced case~\eqref{eq:imbalance_diag_operators}, the fermionic operators~\eqref{eq:ising_diag_operators} now describe Bogoliubov fermionic excitations~\cite{bogoliubov_fermions}. We note again that by using the periodic  bulk energies, our approximation is essentially introducing an intensive contribution coming from the added bonds at the ends of the ladder.

 Finally, the bulk-edge coupling corresponds to a spin-density interaction
\beq
H_{\rm b-e}=V_{\rm v}\left(n_l\tilde{T}_1^x-n_r\tilde{T}_{N-1}^x\right).
\eeq
We can express this interaction in terms of the Bogoliubov excitations, after using the above Jordan-Wigner transformation, together with Eqs.~\eqref{ft_ising} and~\eqref{eq:ising_diag_operators}. In a first approximation, we neglect the Jordan-Wigner string, which allows to derive a bulk-edge coupling that is analogous to the imbalance-induced hybridization~\eqref{eq:hybridization}, namely
\beq
\label{eq:ising_eb}
H_{\rm b-e}=\sum_{n,\alpha} (\tilde{g}^l_{n,\alpha}l^\dagger l^{\phantom{\dagger}}\!\! -\tilde{g}_{n,\alpha}^rr^\dagger r^{\phantom{\dagger}} \!\!)\tilde{\gamma}_\alpha(q_n)+{\rm H.c.}.
\eeq
Once again, in order to obtain the correct bulk-edge couplings $\tilde{g}^\eta_{n,\alpha}$, we have to consider that the connection between the modes of the periodic and open chain require enlarging the number of lattice sites, and thus changing the indexing of the operators in the edge-bulk coupling. This leads to
 \beq
 \begin{split}
 \tilde{g}_{n,+}^l&=+\frac{V_{\rm v}}{\sqrt{2N}}\sin q_n\tilde{u}_{q_n}, \\
 \tilde{g}^l_{n,-}&=-\frac{V_{\rm v}}{\sqrt{2N}}\sin q_n\tilde{v}_{q_n},\\
 \tilde{g}_{n,+}^r&=+\frac{V_{\rm v}}{\sqrt{2N}}\sin \left( q_n(N-1)\right)\tilde{u}_{q_n},\\
  \tilde{g}^r_{n,-}&=-\frac{V_{\rm v}}{\sqrt{2N}}\sin \left( q_n(N-1)\right)\tilde{v}_{q_n}.
 \end{split}
 \eeq 
 
 Accordingly, the Creutz-Hubbard ladder in this formulation $H_{\rm\pi CH }=H_{\rm edge}+H_{\rm bulk}+H_{\rm b-e}$ becomes again similar to a non-interacting two-impurity Fano-Anderson model~\cite{anderson_fano_model}. In this case, the gapless metallic bands in the Fano-Anderson model are substituted by gapped bands of Bogoliubov fermions~\eqref{eq:ising_diag_operators} due to the lack of conservation of the number of Jordan-Wigner fermions. The bulk-edge hybridisation~\eqref{eq:hybridization} is substituted by a bulk-edge coupling~\eqref{eq:ising_eb} that does not conserve the number of Jordan-Wigner fermions either, and depends on the population of the edge modes.
This analogy shall allow us to predict the onset of a topological quantum phase transition in the Creutz ladder caused solely by the Hubbard interactions.

Interpreting once more the dispersive bulk bands as reservoirs, the bulk-edge coupling can have two effects: \emph{(i)} the zero-energy edge modes $\eta=l,r$ may get shifted $\epsilon_\eta=0\to\epsilon_\eta=\Delta\epsilon_\eta$, and dephased with a rate $\tilde{\Gamma}_\eta/2$, \emph{(ii)} the bulk Bogoliubov excitations can mediate an edge-edge interaction (i.e. density-density interactions between the edge fermions mediated by spin-wave-type excitations). In analogy to the imbalance-induced broadening~\eqref{eq:hybridization_spectral_density}, the dephasing rate can be expressed as $\tilde{\Gamma}_\eta= \sum_\alpha\tilde{J}_{\eta\alpha}(\epsilon_\eta)$, where we have introduced the spectral density for the new bulk-edge coupling 
\beq
\tilde{J}_{\eta\alpha}(\omega)=2\pi\sum_n|\tilde{g}_{n,\alpha}|^2\delta(\omega-\alpha\tilde{\epsilon}(q_n)).
\eeq
Accordingly, the dephasing of the superposition states depends on the value of this spectral function at zero energy $\tilde{\Gamma}_\eta=\sum_\alpha\tilde{ J}_{\eta\alpha}(0)$. For $V_{\rm v}/4<2\tilde{t}$, the bulk bands~\eqref{eq:bulk_bands_ising} remain gapped, and thus $J_{\eta\alpha}(0)=0$, such that there is no dephasing. In this regime, there are only level shifts and edge-edge interactions described by 
\beq
H_{\rm e-e}=\Delta\epsilon_{l}n_l+\Delta\epsilon_{r} n_r+U_{\rm ee}n_ln_r,
\eeq
where we have introduced 
\beq
\Delta\epsilon_{\eta}=-\sum_{n,\alpha}\frac{|\tilde{g}^\eta_{n,\alpha}|^2}{\epsilon(q_n)},\hspace{2ex} U_{\rm ee}=\sum_{n,\alpha}\frac{\tilde{g}^l_{n,\alpha}\tilde{g}^r_{n,\alpha}}{\epsilon(q_n)}+{\rm c.c.},
\eeq
Let us note that, in contrast to the imbalanced case, the particle-hole symmetry does not impose the vanishing of the level shifts. Due to the different bulk-edge coupling, these shifts depend on the filling of the Bogoliubov levels for the bulk groundstate $\ket{g_{\rm bulk}}=\tilde{\gamma}_-^{\dagger}(q_1)\tilde{\gamma}_{-}^{\dagger}(q_2)\cdots \tilde{\gamma}_{-}^{\dagger}(q_{N-1})\ket{0_{\rm B}}$, where $\ket{0_{\rm B}}$ is the Bogoliubov vacuum, and thus do not vanish.

If we now use a similar argument as for the imbalance-induced quantum phase transition, we find that \emph{(i)} the bulk-mediated edge-edge interaction has no effect on the degenerate groundstate~\eqref{ni_groundstates}, since both edge modes are not populated simultaneously for a half-filled system. \emph{(ii)} Although the zero-modes are shifted in energy, the topological degeneracy is preserved $\Delta \epsilon_{l}=\Delta\epsilon_{r}$, and no dephasing occurs provided that the Bogoliubov excitations remain gapped. Conversely, when $|\tilde{J}|=2\tilde{t}$, or equivalently $V_{\rm v}=8\tilde{t}$, dephasing takes place signalling that the edge modes are not well-defined single-particle excitations. This argument thus locates the critical point of the interaction-induced quantum phase transition at $\frac{V_{\rm v}}{\tilde{t}}|_{\rm c}=8$, which has been represented by a red circle in the phase diagram of Fig.~\ref{fig_phase_diagram}.

\subsection{Phase diagram of the Creutz-Hubbard ladder}
\label{sec:MPS}

\begin{figure}
\centering
\includegraphics[width=0.9\columnwidth]{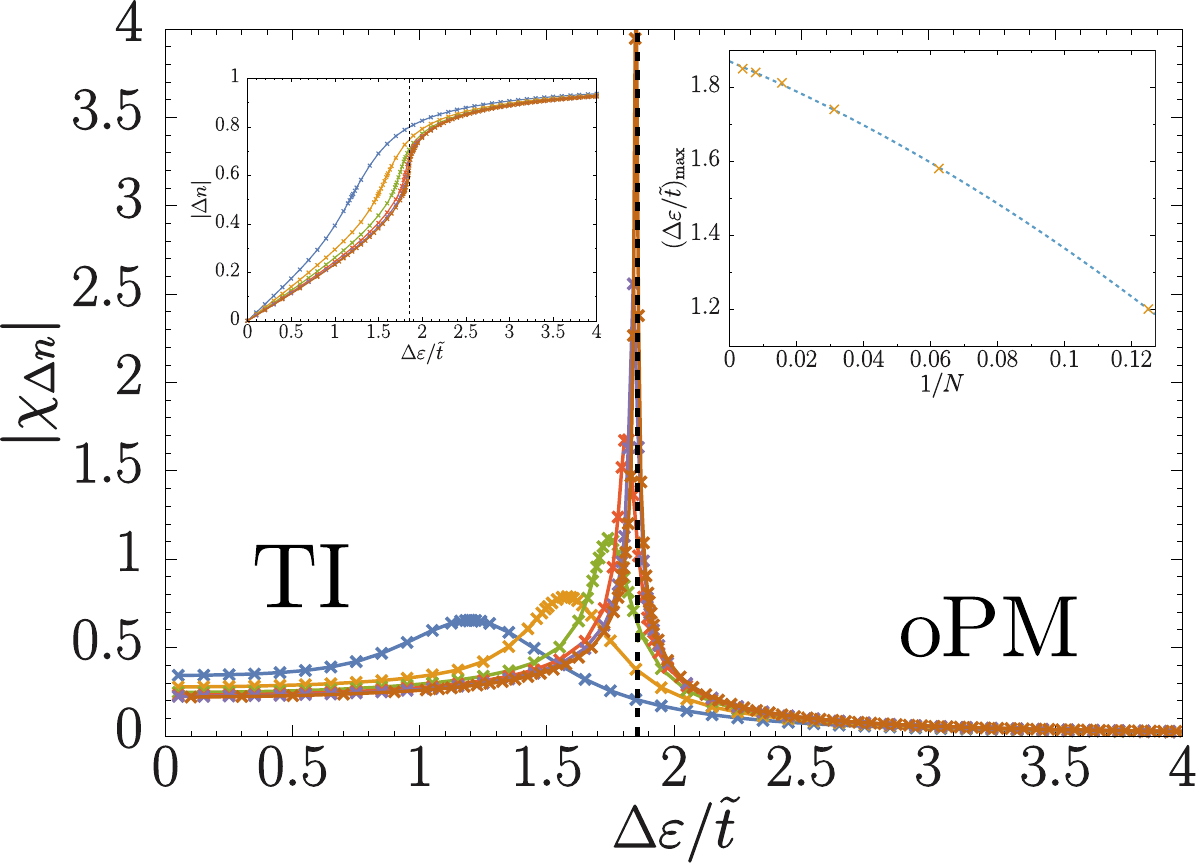}\\
\includegraphics[width=0.9\columnwidth]{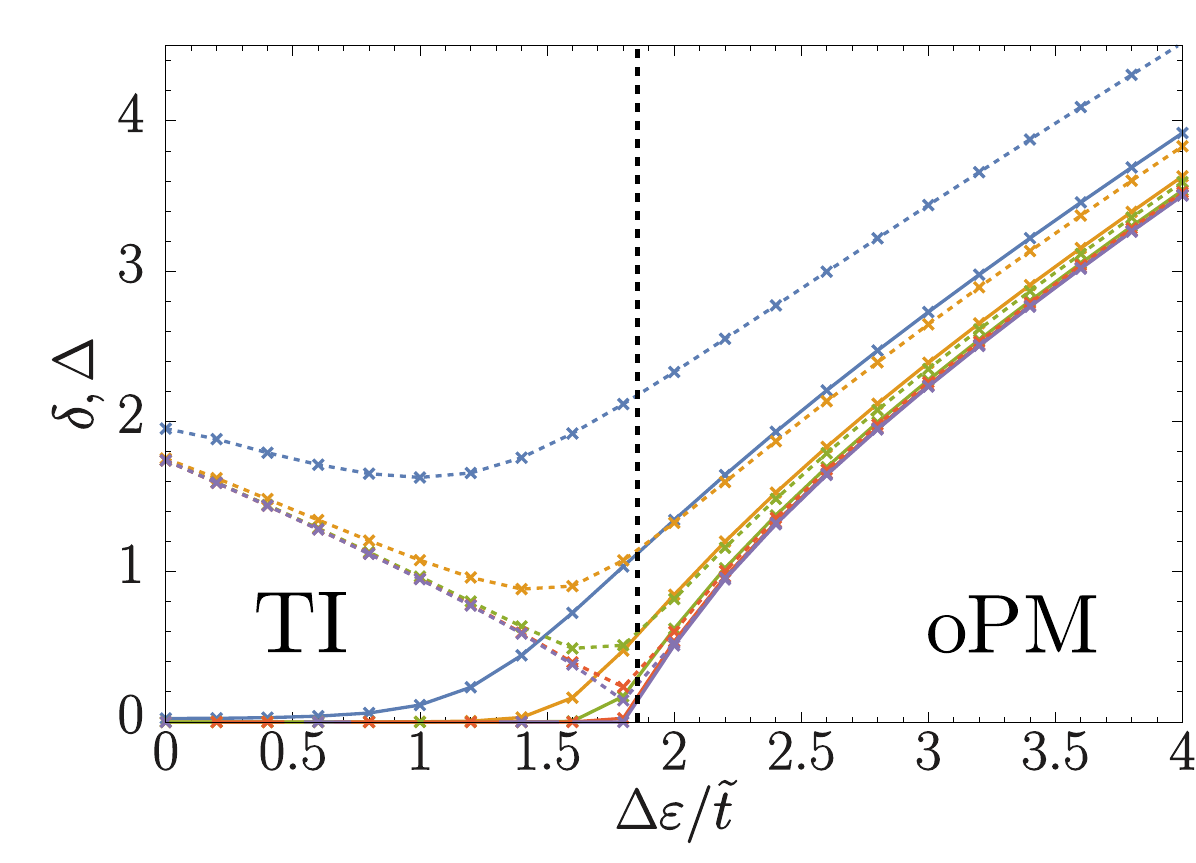}
\caption{ {\bf Paramagnetic susceptibility and energy gaps along the TI-oPM-transition: } Top: The divergence of the magnetization susceptibility $\chi_{\Delta n}$ with growing system size indicates the critical point for a cut through the phase-diagram at $V_v=4.0$ (left inset: occupation imbalance $\Delta n$, right inset: fitted finite-size scaling of the susceptibility maxima assuming up to second-order corrections, $\Delta\varepsilon_{\rm c}(N)=\Delta\varepsilon_{{\rm c}} (1 + aN^{-1} + b N^{-2})$). Bottom: The dashed lines show the finite size results for the energy gap $\Delta$, which is non-zero in both the TI- and the oPM-phase. The quantity $\delta$ (solid lines), on the contrary, is zero in the TI-phase due to the presence of zero energy modes, but achieves a non-vanishing value in the OPM phase. Blue: $N=8$, orange: $N=16$, green: $N=32$, red: $N=64$, violet: $N=128$, brown: $N=256$. The vertical line (black) indicates the transition point ($\Delta\varepsilon_{\rm c} = 1.857$).}
\label{fig:ti-opm}
\end{figure}

Our considerations in the above sections already allowed us to determine the possible phases of the model and their phase boundaries in certain parameter regimes.
In the following, we lay out the full phase-diagram of the model in the $(\frac{\Delta\epsilon}{\tilde{t}},\frac{V_{\rm v}}{\tilde{t}})$ plane using DMRG calculations. 

Our DMRG-code is based on matrix-product states and uses a two-site optimization scheme. It is built on the Abelian Symmetric Tensor Networks Library (developed in collaboration with the group of S. Montangero in Ulm) and capable of implementing multiple Abelian symmetries, such as particle-number conservation. We consider lattices up to $N=256$ sites and virtual bond-dimensions of up to $m=200$.

The analytic and numerical results for the phase diagram, collected in Fig.~\ref{fig_phase_diagram}, are described in the following
subsections:

\subsubsection{Topological insulator to orbital paramagnet phase transition}

As shown in Sec.~\ref{sec:weak_interactions}, the mapping of the Creutz ladder onto a Quantum Ising ladder allows us to predict a critical line~\eqref{eq:critical_line_mean_field} separating the topological insulator (TI) and the orbital paramagnet (oPM) for sufficiently weak interactions. This critical line is represented by a red dashed line that starts form the point $\Delta\epsilon=4\tilde{t}, V_{\rm v}=0$ in Fig.~\ref{fig_phase_diagram}.
 
The Ising transverse magnetization, or equivalently the density imbalance between the legs of the Creutz ladder~\eqref{eq:leg_imbalance}, can serve as a good indicator for this quantum phase transition that can be easily  calculated numerically using the DMRG code. We can determine the critical line by studying the divergence of the imbalance susceptibility $\chi_{\Delta n}=\partial \langle \Delta n\rangle/\partial (\Delta\epsilon/\tilde{t})$ (Fig.~\ref{fig:ti-opm}(top)).

As an alternative means of identifying the TI phase, we can study the behaviour of the ground-state degeneracy in the Creutz-Hubbard model with variable filling. We therefore introduce the single- and two-particle energy gaps
\begin{align}
\label{eq:gaps}
 \Delta&= \lim_{N\rightarrow \infty} \tfrac{1}{2} \left[E(N+2) + E(N-2) - 2 E(N)\right],\\
 \delta&= \lim_{N\rightarrow \infty} \left[E(N+1) + E(N-1) - 2 E(N)\right],
\end{align}
where $E(x)$ is the ground-state energy of a system with $x$ particles. It can be shown that the two quantities coincide for gapless systems ($\Delta=\delta= 0$) and conventional insulators ($\Delta=\delta\neq 0$). In a topological insulator, however, $\delta=0$ due to the presence of zero-energy edge modes while $\Delta\neq 0$ measures (half) the band-gap. In Fig.~\ref{fig:ti-opm}~(bottom), we show that the predicted behaviour is indeed observed.

The critical points obtained through these different observables are represented by yellow stars in the left part of Fig.~\ref{fig_phase_diagram}. As can be seen from these results, the analytical prediction of the phase boundary~\eqref{eq:critical_line_mean_field} is reasonably accurate even for quite large interactions, where the exact critical line given by DMRG departs from a straight line, and bends up.

\begin{figure}
\centering
\includegraphics[width=0.9\columnwidth]{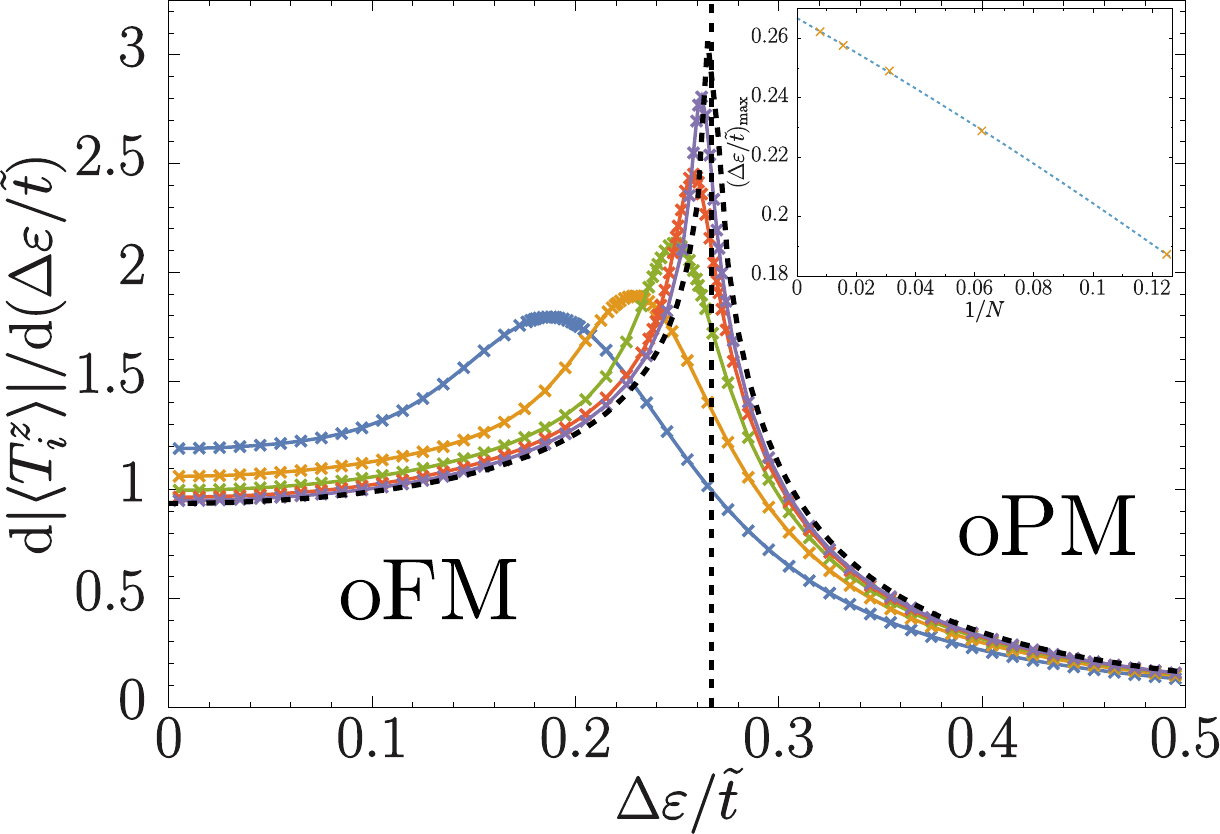}\\
\includegraphics[width=0.9\columnwidth]{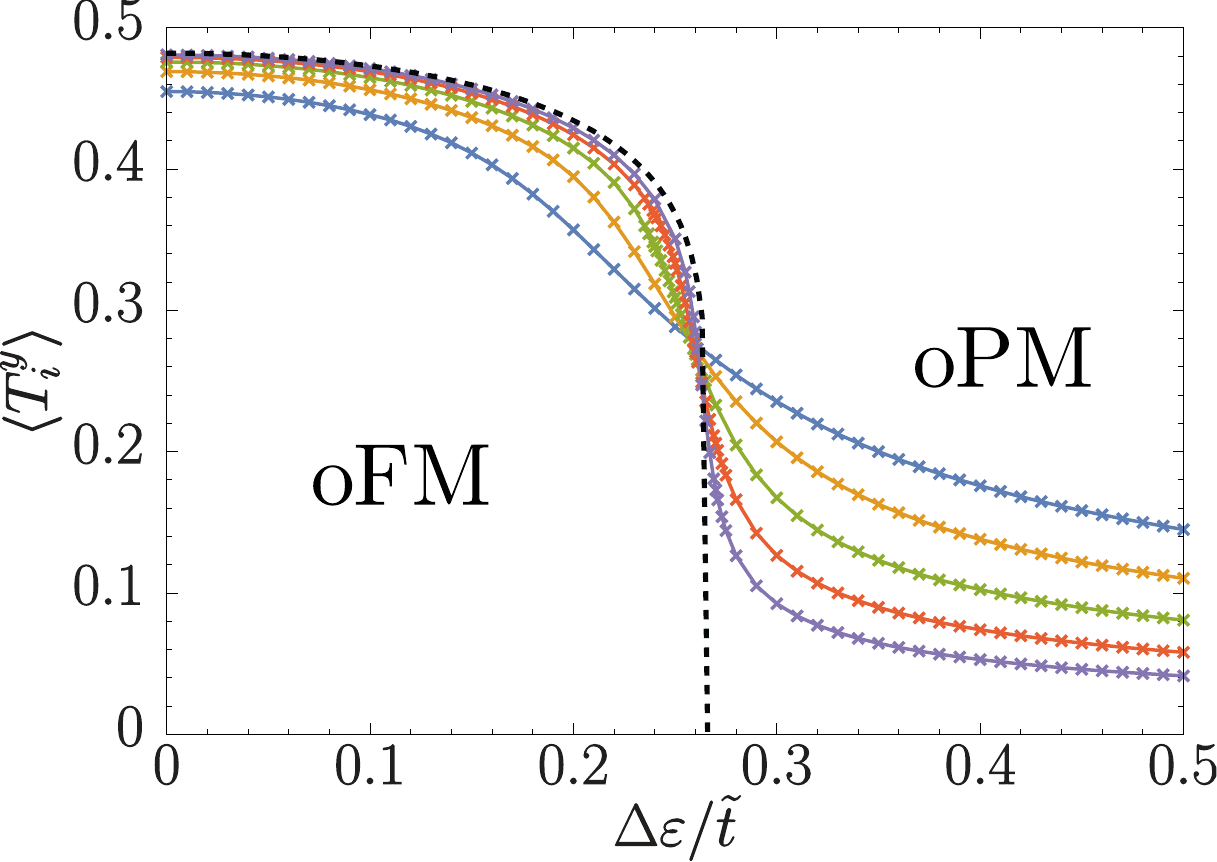}
\caption{ {\bf Ferro- and paramagnetic magnetization along the oFM-oPM-transition: } Top: Paramagnetic magnetization susceptibility for a cut through the phase diagram at $V_v=16\tilde{t}$ and different system sizes (inset: fitted finite-size scaling of the susceptibility maxima assuming up to second-order corrections, $\Delta\varepsilon_{\rm c}(N)=\Delta\varepsilon_{{\rm c}} (1 + aN^{-1} + b N^{-2})$). Bottom: Ferromagnetic magnetization along the same line. Blue: $N=8$, orange: $N=16$, green: $N=32$, red: $N=64$, violet: $N=128$. The light blue vertical line (dashed) in the top figure indicates the critical point (here: $\Delta\varepsilon_{\rm c}/\tilde{t}=0.266$). The black dashed curves indicate the analytical predictions of an Ising model with the same critical point (and a saturation of the ferromagnetic magnetization $\langle T_i^y\rangle_{\rm max}=0.48$). The effective Ising model in Eq.~\eqref{si_spins} suggests $\Delta\varepsilon_{\rm c}/\tilde{t}=0.25$ and $\langle T_i^y\rangle_{\rm max}=0.5$. \label{fig:ofm-opm}}
\end{figure}

\subsubsection{Orbital ferromagnet to orbital paramagnet phase transition}

In Sec.~\ref{sec:XYmodel}, we introduced an effective orbital Ising model in the limit of very strong interactions, which allowed us to predict a critical line~\eqref{eq:phase_boundary_Ising} separating the 
orbital ferromagnet (oFM) and orbital paramagnet (oPM). This critical line is represented by a yellow dashed line in Fig.~\ref{fig_phase_diagram}.

Indeed by measuring the paramagnetic and ferromagnetic magnetization ($\langle T_i^z\rangle$ resp. $\langle T_i^y\rangle$ in Eq.~\eqref{si_spins}), we confirm that these quantities scale equally, and identify the phase-transition point also for finite interactions (Fig.~\ref{fig:ofm-opm}). Technically, we determine the paramagnetic magnetization by measuring the fermionic observable that defines $T_i^z$, which is proportional to the leg density imbalance discussed above (see Eq.~\eqref{si_spins}). In order to avoid problems due to incomplete symmetry breaking when studying the ferromagnetic order-parameter $\langle T_i^y\rangle$ (i.e. between the possible alignments in the ferromagnetic phase), we determine instead the zero-momentum component of the orbital magnetic structure factor
\begin{align}
 S_{T_y T_y}(k) = \frac{1}{N^2} \sum_{i,j}\ee^{\ii k(i-j)} \left\langle T_i^y T_j^y \right\rangle,
\end{align}
which yields the desired ferromagnetic magnetization in the thermodynamic limit $\langle T_i^y\rangle=( S_{T_y T_y}(0))^{1/2}$.

We observe for both quantities an Ising-like scaling, which differs from the strong-coupling prediction only by a renormalization of the critical point and of the maximum ferromagnetic magnetization (comp. Fig.~\ref{fig:ofm-opm}).

The critical points obtained through these magnetizations are represented by yellow stars in the right part of Fig.~\ref{fig_phase_diagram}. As can be seen from these results, the analytical prediction of the phase boundary~\eqref{eq:critical_line_mean_field} is reasonably accurate even for moderate interactions. 

\subsubsection{Topological insulator to orbital ferromagnet phase transition} 
\begin{figure}
\centering
\includegraphics[width=0.9\columnwidth]{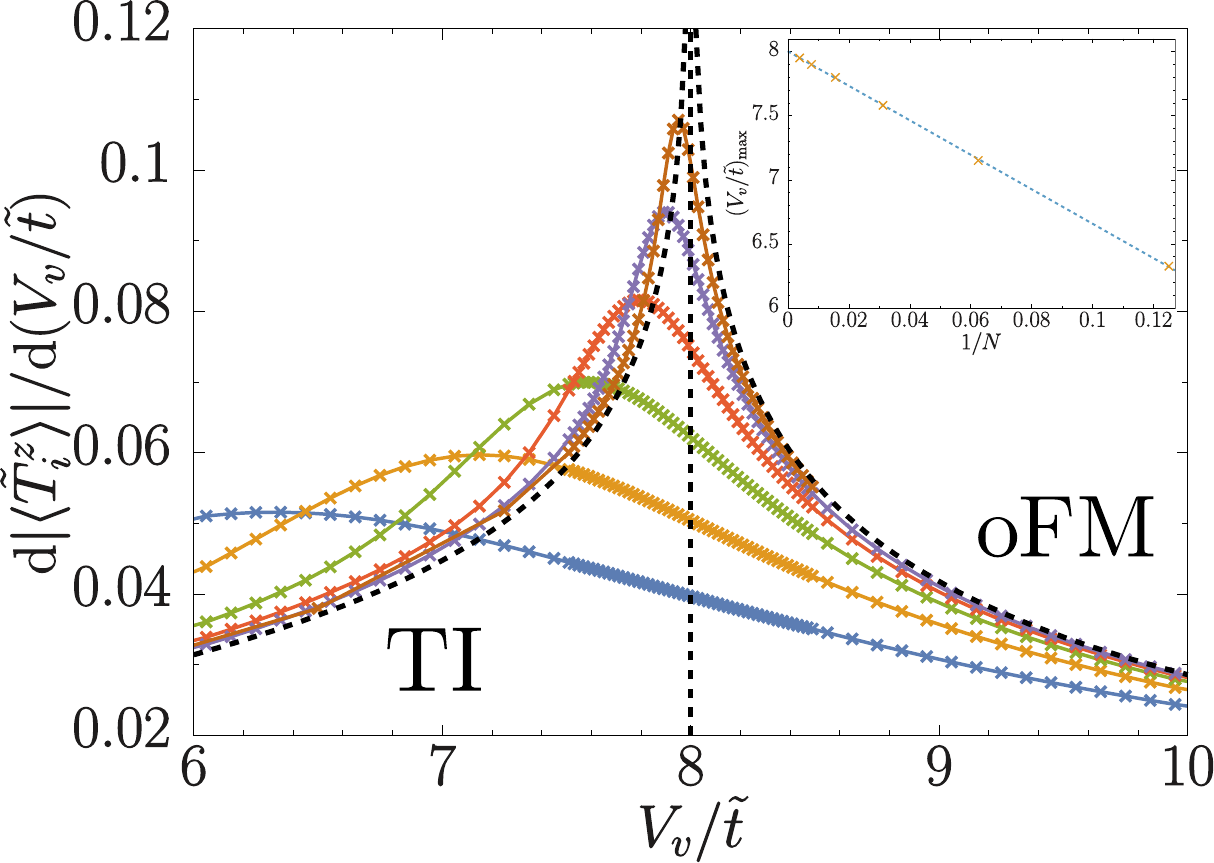}
\caption{ {\bf Paramagnetic magnetization susceptibility along the the TI-oFM-transition:} Blue: $N=8$, orange: $N=16$, green: $N=32$, red: $N=64$, violet: $N=128$, brown: $N=256$. The black dashed lines indicate the TD-result in a TFIM-model. Inset: The finite-size scaling of the maxima of the susceptibility yields $(V_v/\tilde{t})_{\rm c, num} = 8.003$ (here fitted via $V_{v,\rm c, num}(N)=V_{v,{\rm c,num}} (1 + aN^{-1})$), in good agreement with the analytical result $(V_v/\tilde{t})_{\rm c} = 8$ .} \label{fig:ti-ofm}
\end{figure}

In Sec.~\ref{sec:int_regime}, we derived an interesting connection between the balanced Creutz-Hubbard ladder for intermediate interactions and a Fano-Anderson-type model, which allowed us to 
predict the extension of the topological phase along the $(0,\frac{V_{\rm v}}{\tilde{t}})$ axis of the phase diagram until a critical point $\frac{V_{\rm v}}{\tilde{t}}=8$. Beyond this point, the long-range ordered orbital Ising magnet sets in, and the topological edge modes disappear into the bulk. This critical point is represented by a red circle in Fig.~\ref{fig_phase_diagram}.

The numerical analysis (Fig.~\ref{fig:ti-ofm}) confirms the validity of the effective Ising-model derived in Eq.~\eqref{eq:eff_ising}, and the exact location of this critical point. Moreover, in the case of finite imbalance ($\Delta \varepsilon \neq 0$), the divergence of the paramagnetic susceptibility serves as a criterion for the determination of the phase-boundary (see inset Fig.~\ref{fig:ti-ofm}).
The critical points obtained by these means are represented by yellow stars in the middle part of Fig.~\ref{fig_phase_diagram}.

\subsubsection{Conformal field theories for the critical lines and entanglement spectrum for the phases}

So far, we have used a conventional condensed-matter approach to explore numerically the phase diagram of the model, which is based on exploiting energy gaps, susceptibilities, and correlation functions to identify phases with long-range order or symmetry-protected topological phases, and critical lines that separate them. An alternative approach, based on the groundstate entanglement, has recently become a complementary method to understand the phase diagram of quantum many-body models~\cite{entanglement_rmp}. For instance, the pair-wise concurrence 'susceptibility' can serve as a probe to localize quantum critical points~\cite{concurrence_phase_transition}, displaying a scaling behavior similar to that of certain observables in the more conventional condensed-matter approach. Other entanglement measures can also serve as probes of quantum criticality, as occurs for the block entanglement entropy $S(l)=-{\rm Tr}\{\rho_l\log\rho_l\}$, where $\rho_l={\rm Tr}_{L-l}\{\ket{\epsilon_{\rm gs}}\bra{\epsilon_{\rm gs}}\}$ is the reduced density matrix of the left block with $l$ sites for a bipartition of a chain of $L$ sites. Remarkably enough, not only does the block entanglement entropy serve as a probe of criticality due to its divergence at a phase transition, but its scaling with the system size also reveals the central charge $c$ of the conformal field theory (CFT) underlying the critical behavior~\cite{ent_entropy_critical,entanglement_scaling}. For a critical system with open boundary conditions, the block entanglement entropy scales as
 \beq
 S(l) = \frac{c}{6}\ln\left(\frac{2L}{\pi}\sin\frac{\pi l}{L}\right) + a,
 \eeq
 where we have introduced a non-universal constant $a$. Since such entanglement entropy can be easily recovered from our MPS numerical results, calculating the central charge of the different critical lines of our phase diagram can serve as an additional confirmation of our previous derivations.

In Sec.~\ref{sec:weak_interactions}, we argued that the synthetic Creutz ladder for sufficiently weak interactions can be understood as a couple of Ising models of length $L=N$ with a renormalized transverse field. Accordingly, the corresponding CFT should have central charge of $c=1/2+1/2=1$, such that we would expect the scaling $ S(l) = \frac{1}{6}\ln\left(\frac{2N}{\pi}\sin\frac{\pi l}{N}\right) + a$. This is a natural connection to the non-interacting regime of Sec.~\ref{sec:int_tqpt}, where we argued that the phase transition can be understood in terms of a Dirac fermion with a Wilson mass in lattice gauge theories. This Wilson fermion becomes massless at the critical point, and corresponds to the CFT of a single massless Dirac fermion with central charge $c=1$.

For the strongly-interacting regime of Sec.~\ref{sec:XYmodel}, we showed that the oFM-oPM quantum phase transition can be predicted in terms of a single Ising model of length $L=N$ in a transverse field. Accordingly, the corresponding CFT should have central charge of $c=1/2$, and $ S(l) = \frac{1}{12}\ln\left(\frac{2N}{\pi}\sin\frac{\pi l}{N}\right) + \tilde{a}$. In contrast to the previous case, the critical phenomena is governed by the CFT of a single Majorana fermion with central charge $c=1/2$.

Finally, in the intermediate interacting regime of Sec.~\ref{sec:int_regime}, we argued that the relevant physics to understand the TI-oFM phase transition is by approximating a complicated non-standard Hubbard model with a simpler quantum impurity model for the edges coupled to a bulk of spins described by yet another Ising model of $L=N-1$ sites in a transverse field. Technically, however, we determine the entanglement entropy between the $N$ physical sites in the original basis and therefore measure $ S(l) = \frac{1}{12}\ln\left(\frac{2N}{\pi}\sin\frac{\pi l}{N}\right) + a'$ for a system of length $L=N$, corresponding to the CFT of a single Majorana fermion with a central charge of $c=1/2$.

\begin{figure}
\centering
\includegraphics[width=0.9\columnwidth]{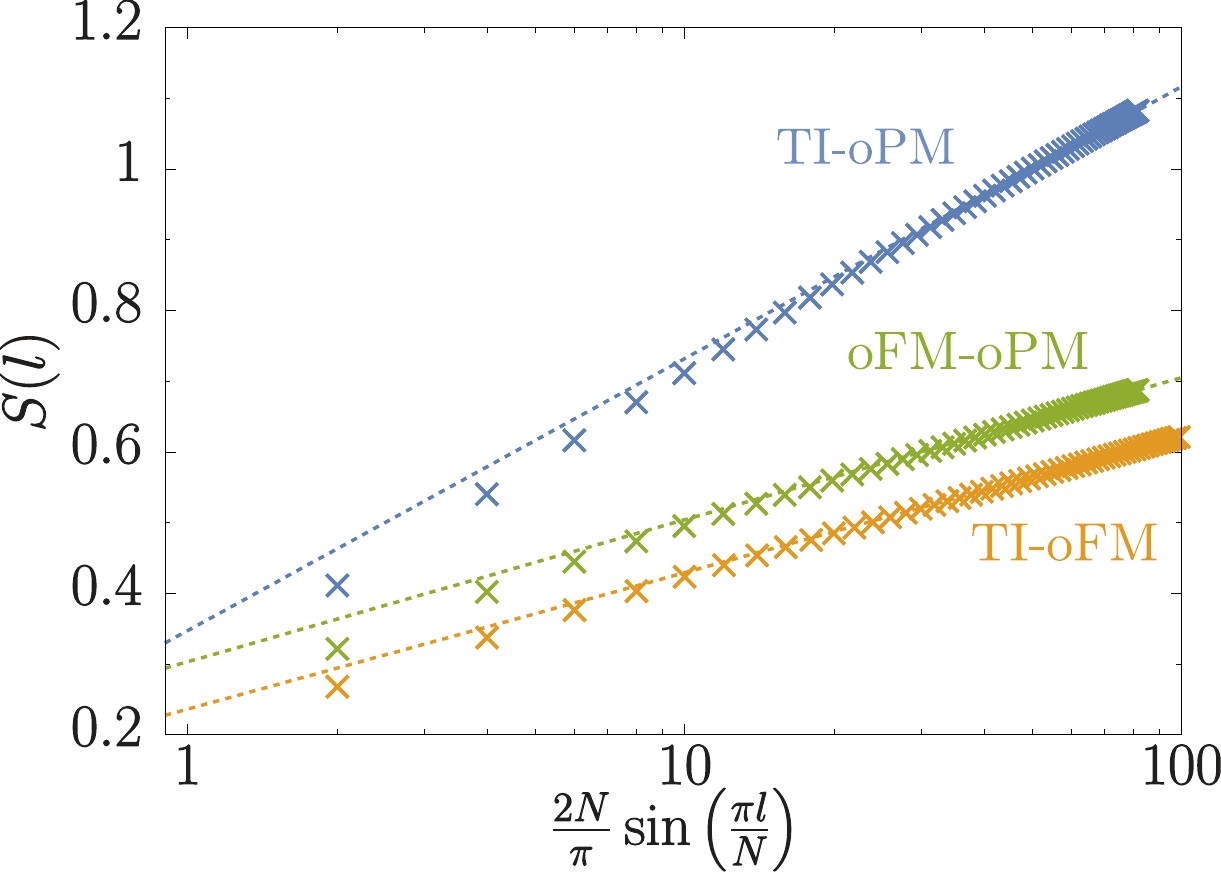}
\caption{ {\bf Scaling of the entanglement entropy for critical points on the different transition lines.} The prefactor $c$ in $S(l) = \tfrac{c}{6}\ln(\tfrac{2N}{\pi}\sin\tfrac{\pi l}{N}) + {\rm const.}$ identifies the central charge of a critical phase. Here we show the entanglement entropy in systems with $N$ sites and fit the data for $N/4<l\le N/2$ . The fitting results for the TI-oPM transition  yield $c=1.003$  for $V_v/\tilde{t}=4.0,\Delta\varepsilon/\tilde{t}=1.857, N=128$ (blue line), the oFM-oPM transition yield $c=0.524$ for  $V_v/\tilde{t}=16,\Delta\varepsilon=0.266, N=128$ (green line), and the TI-oFM transition  yield $c=0.503$ for $V_v/\tilde{t}=0,\Delta\varepsilon/\tilde{t}=8, N=256$ (yellow line). All these numerical fits  agree considerably well with the model predictions $c\in\{1,1/2,1/2\}$.} \label{fig:entscal}
\end{figure}

We confirm the above predictions through the numerical determination of the central charge along the critical lines in three representative cases (see Fig.~\ref{fig:entscal}). We find central charge values agreeing with $c=1/2$ for the TI-oFM and the oFM-oPM-transition. The charge $c=1$ along the TI-oPM-transition originates from the hybrid nature of the Ising-model describing it (see Eq.~\eqref{eq:two_copy_ising}). Building on these results, we depict the central charges of the three critical lines of our phase diagram in Fig.~\ref{fig_phase_diagram}. Interestingly, the massless Dirac fermion $c=1$ governing the topological phase transition Ti-oPM for sufficiently weak interactions, is split into a pair of massless Majorana fermions $c=\half +\half$ at the tri-critical point, each of which governs the critical properties of the TI-oFM and oFM-OPM quantum phase transitions, respectively. At this tri-critical point, the central charge is conserved, and the two Majorana fermions paired to yield the Dirac fermion governing the TI-oPM phase transition, become un-paired and describe individually the critical properties of the two other phase transitions of the model.

\begin{figure}
\centering
\includegraphics[width=0.9\columnwidth]{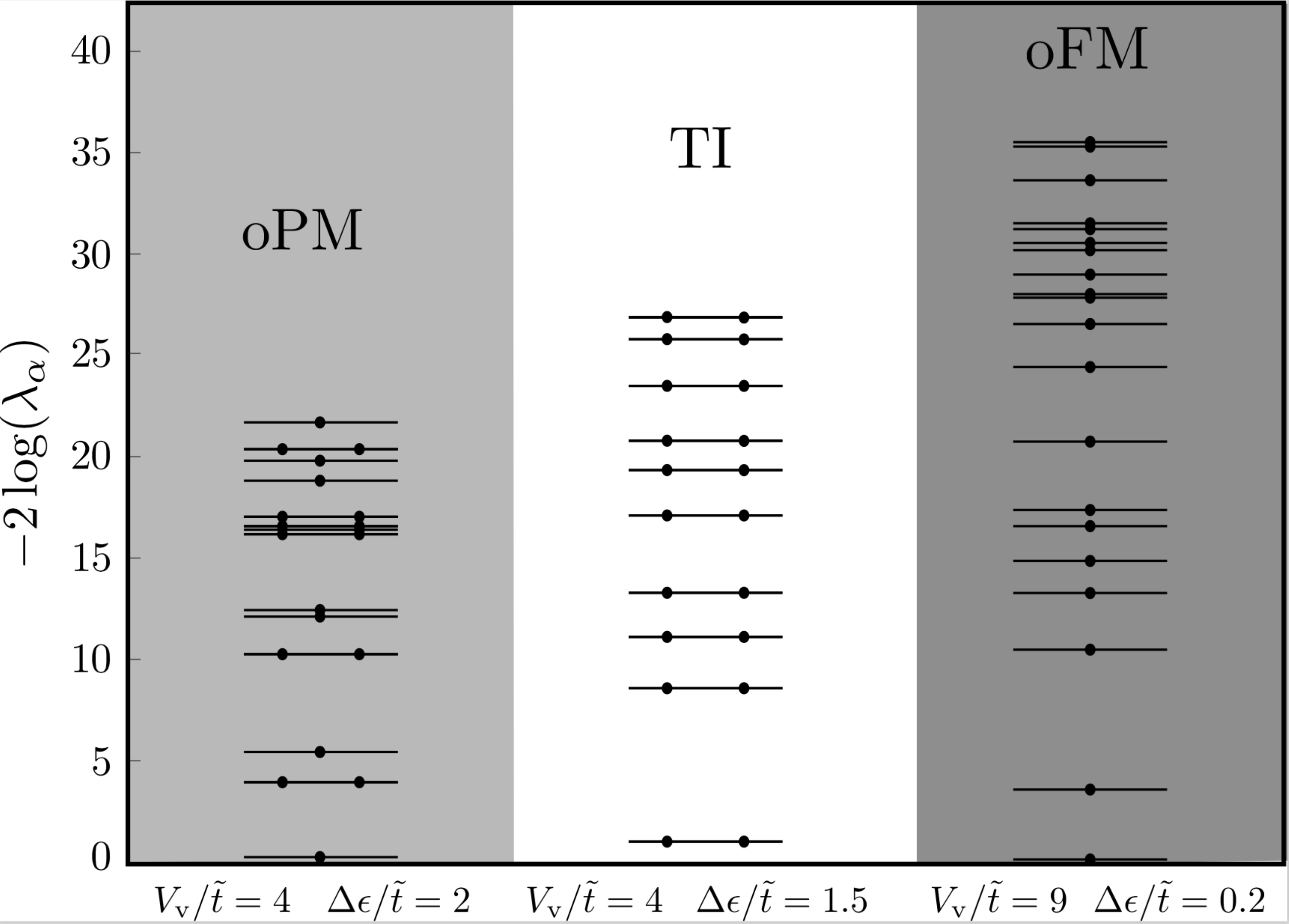}
\caption{ {\bf Degeneracies of the entanglement spectrum for different phases.} For a ladder of length $L=128$ and for a bipartition in the half chain, the twenty lower eigenvalues of the ES are depicted for the three different phases. The dots represent the degeneracy of the corresponding eigenvalue. In the TI phase, for $V_{\rm v}/\tilde{t} = 4$ and $\Delta_{\epsilon}/\tilde{t}=1.5$ the eigenvalues are all doubly degenerate. In the oPM phase, for $V_{\rm v}/\tilde{t} = 4$ and $\Delta_{\epsilon}/\tilde{t}=2$, and in the oFM phase for $V_{\rm v}/\tilde{t} = 9$ and $\Delta_{\epsilon}/\tilde{t}=0.2$ almost all the eigenvalues are not degenerate. We found the same behaviour elsewhere in  phase space.} \label{fig:entspectrum}
\end{figure}

In the previous sections, we provided several indicators of the non-trivial topological nature of the TI phase, such as  the resilience of the zero-energy edge modes as interactions are switched on (see Sec.~\ref{sec:int_regime}), or the different single- and two-particle gaps~\eqref{eq:gaps} that distinguish  topological and non-topological phases (see Sec.~\ref{sec:MPS}). Let us now provide further evidence by using entanglement properties of the groundstate.  In particular, a strong signature of the presence of topological order can be extracted from the study of the \textit{entanglement spectrum} (ES) \cite{li_haldane_es}. Similarly to the case of the entanglement entropy, to define the ES we take a bipartition of the ladder and the resulting Schmidt decomposition of the ground state $\ket{\psi} = \sum_i \lambda_i\ket{\psi_L^i}\otimes\ket{\psi_{L-l}^i}$, where $\ket{\psi_L^i}$ and $\ket{\psi_{L-l}^i}$ are basis vectors of the two parts, satisfying the orthogonality condition $\langle \psi_l^j \ket{\psi_{L-l}^i} = \delta_{ij}$, whereas the $\lambda_i$ are the Schmidt values. By convention, the ES is defined as a logarithmic rescaling of the Schmidt values, $-2\log(\lambda_i)$ and it can be again extracted  straightforwardly from MPS calculations. As originally pointed out in \cite{pollmann_es} in the context of the characterization of the Haldane phase of Heisenberg-type magnets, the degeneracy of the ES robustly identifies the symmetries protecting the topological phase. As shown in Fig. \ref{fig:entspectrum} in the imbalanced Creutz-Hubbard model, the ES in the TI phase clearly shows  doubly-degenerate eigenvalues, whereas in the oPM and oFM phases the ES is trivial and almost completely non-degenerate. This  supports the  topological nature of the wide region of the phase diagram labelled as TI (see Fig.~\ref{fig_phase_diagram}), and demonstrates that the topological insulating phase survives to considerably strong interactions.

\section{ Synthetic Creutz-Hubbard ladder}
\label{sec:proposal}

In this section, we discuss an experimental setup capable of implementing the \emph{imbalanced Creutz-Hubbard Hamiltonian}~\eqref{eq:pi_ch}. We shall focus on a particular set of microscopic couplings that can be realised with ultracold atoms in optical lattices by exploiting the tools of laser-assisted tunnelling.

\subsection{Cold-atom Creutz-Hubbard ladder}
\label{sec:schl}
\begin{figure*}
\centering
\includegraphics[width=1.9\columnwidth]{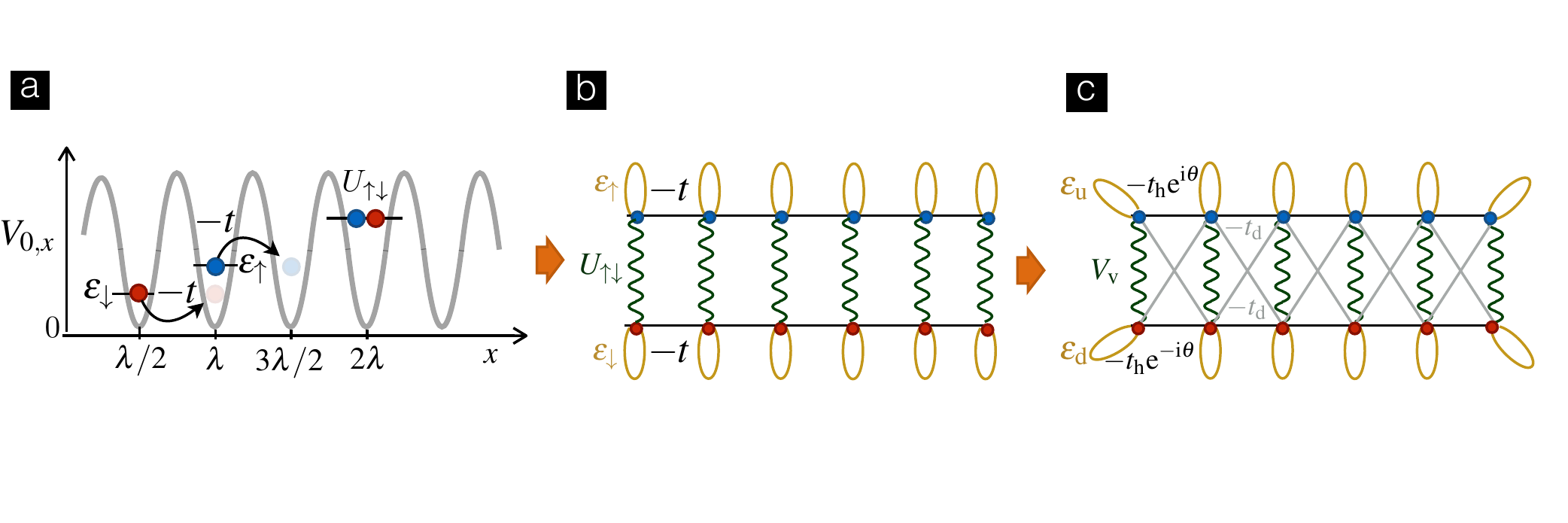}
\caption{ {\bf Ladder scheme for a binary hyperfine mixture of ultracold atoms in an optical lattice: } {\bf (a)} Atoms in two hyperfine states $\ket{{\uparrow}}=\ket{F=I-\half,M}$ (blue circles), $\ket{{\downarrow}}=\ket{F'=I+\half,M'}$ (red circles) are trapped at the minima of an optical lattice. At low temperatures, the kinetic energy of the atoms can be described as tunnelling of strength $-t$ between the lowest energy levels $\epsilon_{\uparrow},\epsilon_{\downarrow}$ of neighbouring potential wells. Additionally, the s-wave scattering of the atoms leads to contact interactions of strength $U_{\uparrow\downarrow}$ whenever two fermionic atoms of different internal state meet on the same potential well. {\bf (b)} Ladder representation of the binary mixture, where each hyperfine state corresponds to a different leg of the ladder. The tunnelling (on-site energy) is represented by solid lines linking neighbouring (same) sites within each leg, whereas the on-site interactions are represented by curly 
lines among two neighbouring sites that belong to different legs. {\bf (c)} Creutz-Hubbard model obtained by including a Peierls phase in the intra-leg tunnelings $-t\to\pm-t_{\rm h}\ee^{\pm\ii\theta}$, and introducing inter-leg tunnelings $-t_{\rm d}$ represented by solid lines along diagonal rungs of the the ladder.}
\label{fig_ladder_scheme}
\end{figure*}

We consider a cubic state-independent optical lattice that traps a two-component atomic gas with hyperfine states $\ket{{\uparrow}}=\ket{F=I-\half,M}, \ket{{\downarrow}}=\ket{F'=I+\half,M'}$ (see Fig.~\ref{fig_ladder_scheme}{\bf (a)}). In the Wannier basis, the Hamiltonian corresponds to the standard Hubbard model~\cite{fermi_hubbard_ol}, namely
\beq
\label{eq:hubbard}
H=\sum_{i=1}^N\sum_{\sigma=\uparrow,\downarrow}\!\!\!\left(-tf_{i+1,\sigma}^{\dagger}f_{i,\sigma}^{\phantom{\dagger}}+\frac{\epsilon_\sigma }{2}n_{i,\sigma}+{\rm H.c.}\right)+\sum_iU_{\uparrow\downarrow}n_{i,\uparrow} n_{i,\downarrow},
\eeq
where the fermionic operators $ f_{i,\sigma}^{{\dagger}},f_{i,\sigma}^{\phantom{\dagger}}$ create and annihilate an atom in the electronic state $\sigma$ at site $i$ of the lattice, and $n_{i,\sigma}=f_{i,\sigma}^{{\dagger}}f_{i,\sigma}^{\phantom{\dagger}}$ are the number operators. In this equation, $t$ is the tunnelling strength of atoms between neighbouring potential wells along the $x$-axis, $\epsilon_\sigma$ stand for the energies of the electronic levels, and $U_{\uparrow\downarrow}$ corresponds to the on-site interaction strength due to s-wave scattering. As customary~\cite{bose_hubbard_ol}, tunnelings and interactions of a longer range, as well as tunnelings along the $y$ and $z$ axes, have been neglected in this expression. This is justified for sufficiently deep optical lattices $V_{0,y},V_{0,z}\gg V_{0,x}\gg E_{\rm R}$, where $V_{0\alpha}$ are the corresponding amplitudes of the optical potential $V(\boldsymbol{r})=\sum_{\alpha=x,y,z}V_{0\alpha}\sin^2(kr_{\alpha})$, $E_{\rm R}=k^2/2m$ is the recoil energy, and $k$ is the wavevector of the retro-reflected laser beams forming the optical lattice. In this regime, the parameters of Eq.~\eqref{eq:hubbard}
can be expressed as 
\beq 
\label{eq:tU}
t=\frac{4E_{\rm R}}{\sqrt{\pi}}\!\left(\frac{V_{0,x}}{E_{\rm R}}\right)^{\!\!\frac{3}{4}}\!\!\ee^{-2\sqrt{\frac{V_{0x}}{E_{\rm R}}}},\hspace{1ex} U_{\uparrow\downarrow}=\sqrt{\frac{8}{\pi}}ka_{\uparrow\downarrow}E_{\rm R}\!\left(\!\frac{ V_{0x}V_{0y}V_{0z}}{E_{\rm R}}\!\right)^{\!\!\fourth},
\eeq
where $a_{\uparrow\downarrow}$ is the s-wave scattering length for the collision of two atoms in the two different hyperfine states.

The first step towards a possible implementation of the Creutz-Hubbard model based on Eq.~\eqref{eq:hubbard} is to represent the legs of the ladder by the two hyperfine states~\cite{creutz_kz,qft_ti_ol} (see Fig.~\ref{fig_ladder_scheme}{\bf (b)}). This might be interpreted as the smallest-possible synthetic dimension along which real~\cite{synthetic_dimension} or complex~\cite{gauge_fields_synthetic} tunnelings can be implemented via Raman transitions, leading to recent experiments realising synthetic gauge fields~\cite{gauge_fields_sd,gauge_fields_sd_II}. Unfortunately, 
the Creutz ladder involves more complicated tunnelings (see Fig.~\ref{fig_ladder_scheme}{\bf (c)}) that cannot be directly obtained using this scheme. One possibility to implement the required tunnelings could be to combine bi-chromatic optical lattices with additional Raman transitions and a staggered optical potential, which allows for a very flexible toolbox to realise spin-dependent tunnelings~\cite{qft_ti_ol}. However, in view of the success of experimental schemes based on periodic modulation of the lattice~\cite{gauge_ol,haldane_ol} that can be understood in terms of Floquet engineering~\cite{floquet_atoms}, we hereby present an alternative scheme to achieve the desired Hamiltonian combining periodic modulations with a generalisation of the Raman-assisted scheme~\cite{jz_gauge_fields} applied to a spin-independent optical lattice.

\begin{figure*}
\centering
\includegraphics[width=1.7\columnwidth]{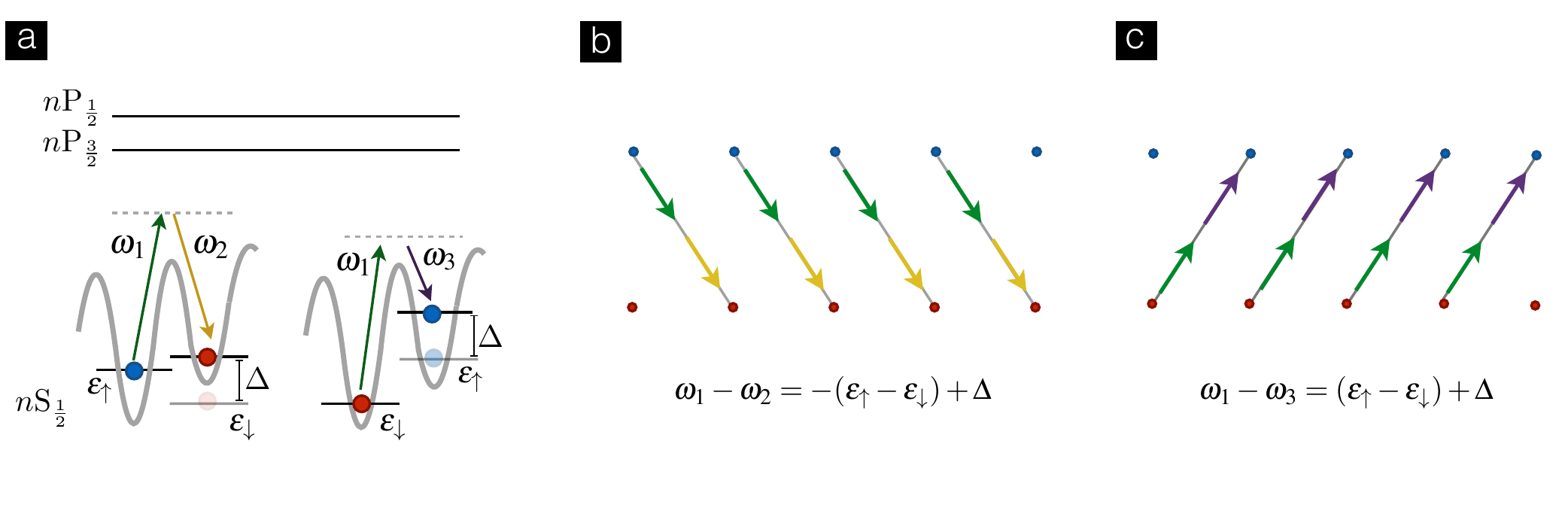}
\caption{ {\bf Raman-assisted tunnelling scheme for a binary hyperfine mixture of ultracold atoms in an optical lattice: } {\bf (a)} Atoms in a given hyperfine state from the groundstate manifold $nS_{1/2}$ can tunnel to the neighbouring potential well changing its hyperfine state by means of a Raman transition that virtually populates an excited state from the manifolds $nP_{1/2}, nP_{3/2}$. {\bf (b)} The inter-leg tunnelling of $\ket{{\uparrow}}$ (blue circles) to $\ket{{\downarrow}}$ (red circles) is mediated by a pair of lasers providing the energy necessary to overcome the offset $\omega_1-\omega_2=(\epsilon_{\downarrow}-\epsilon_{\uparrow})+\Delta$, and a recoil kick to assist the tunnelling. {\bf (c)} The inter-leg tunnelling of $\ket{{\downarrow}}$ (red circles) to $\ket{{\uparrow}}$ (blue circles) is mediated by a pair of lasers providing the energy necessary to overcome the offset $\omega_1-\omega_3=(\epsilon_{\uparrow}-\epsilon_{\downarrow})+\Delta$, and a recoil kick to assist the 
tunnelling.}
\label{fig_raman_scheme}
\end{figure*}

\subsubsection{ State-independent energy gradient}

 We consider a linear energy gradient that tilts the optical lattice independently of the hyperfine state. Such a linear gradient can be obtained by accelerating the lattice, or by exploiting the ac-Stark effect, and yields
\beq
\label{eq:tilt}
H_{\rm tilt}=\sum_{i,\sigma} \Delta i \hspace{0.2ex}f_{i,\sigma}^{\dagger} f_{i,\sigma}^{\phantom{\dagger}}.
\eeq
Note that we have considered the regime $\Delta\ll V_{0x}$, such that the gradient does not modify the inter-site terms of the original Hubbard model~\eqref{eq:hubbard}, but only leads to the local on-site term~\eqref{eq:tilt} in the Wannier basis. Moreover, we impose $t\ll\Delta$, such that the tilt inhibits the original intra-leg tunnelling in Fig.~\ref{fig_ladder_scheme}{\bf (b)}. The goal now is to re-activate the tunneling against this gradient, by generalising the ideas of schemes based on Raman transitions~\cite{jz_gauge_fields}, or on periodic modulations~\cite{gauge_fields_periodic_modulations} with additional shallower optical lattices~\cite{gauge_ol}.

\subsubsection{ Raman-assisted tunnelling}

 We consider three additional laser beams. These lead to a couple of Raman two-photon excitations that assist the inter-leg tunnelings (see Fig.~\ref{fig_raman_scheme}{\bf (a)}). A first pair of laser beams with frequencies $\omega_1,\omega_2$ assists the inter-leg tunnelings in Fig.~\ref{fig_raman_scheme}{\bf (b)} by virtually populating an excited state. The effect of the lasers is two-fold, they provide the energy to overcome the potential offset in the tunnelling process $\omega_1-\omega_2=(\epsilon_{\downarrow}-\epsilon_{\uparrow})+\Delta$, and they also induce a recoil kick $\delta k=({\bf k}_1-{\bf k}_2)\cdot{\bf e}_x$ along the $x$-axis, thus allowing the tunnelling to occur (i.e. overlap of neighbouring Wannier functions). The other pair of laser beams assists the inter-leg tunnelling in Fig.~\ref{fig_raman_scheme}{\bf (c)} provided that $\omega_1-\omega_3=(\epsilon_{\uparrow}-\epsilon_{\downarrow})+\Delta$, also yielding a recoil kick $\delta \tilde{k}=({\bf k}_1-{\bf k}_3)\cdot{\bf e}_x$. 
Altogether, these Raman-assisted terms lead to
\beq
\label{Raman}
H_{\rm Raman}=\sum_{i} \frac{1}{2}\left({\Omega}f_{i+1,\downarrow}^{\dagger} f_{i,\uparrow}^{\phantom{\dagger}}+\tilde{\Omega}f_{i+1,\uparrow}^{\dagger} f_{i,\downarrow}^{\phantom{\dagger}}\right) +{\rm H.c.},
\eeq
where we have introduced the Raman-assisted tunnelling strengths 
\beq
\label{eq:OmegaR}
\Omega=\Omega_{12}\ee^{-\frac{\pi}{4}\sqrt{\frac{V_{0x}}{E_{\rm R}}}},\hspace{2ex}\tilde{\Omega}=\Omega_{13}\ee^{-\frac{\pi}{4}\sqrt{\frac{V_{0x}}{E_{\rm R}}}},
\eeq
which are expressed in terms of the respective two-photon Rabi frequencies $\Omega_{12}, \Omega_{13}$ corresponding to the particular Raman transition, and an exponential term due to the laser-assisted overlap of the neighbouring Wannier orbitals, which has been calculated using a Gaussian approximation. We shall consider the limit where the Raman-assisted tunneling strengths are set to be equal $\Omega=\tilde{\Omega}$. To arrive at these expressions, we have considered again the regime of deep optical lattices, and that $\delta k\cdot \lambda=2\pi n$, $\delta \tilde{k}\cdot \lambda=2\pi \tilde{n}$, where $n,\tilde{n}\in\mathbb{Z}$ and $\lambda$ is the wavelength of the retro-reflected laser beam that leads to the original optical lattice. This can be achieved by controlling the angle of the Raman lasers with respect to the $x$-axis. More importantly, we have assumed that the two-photon Rabi frequencies fulfil $|\Omega_{12}|,|\Omega_{13}|\ll\Delta$, such that on-site Raman transitions are highly off-resonant, and can be neglected. 

Let us note that, by avoiding a state-dependent optical lattice that traps each of the two components differently, our scheme is not subjected to the heating problems that become specially troublesome for fermionic atoms~\cite{gd_gauge_fields}.

\begin{figure*}
\centering
\includegraphics[width=1.7\columnwidth]{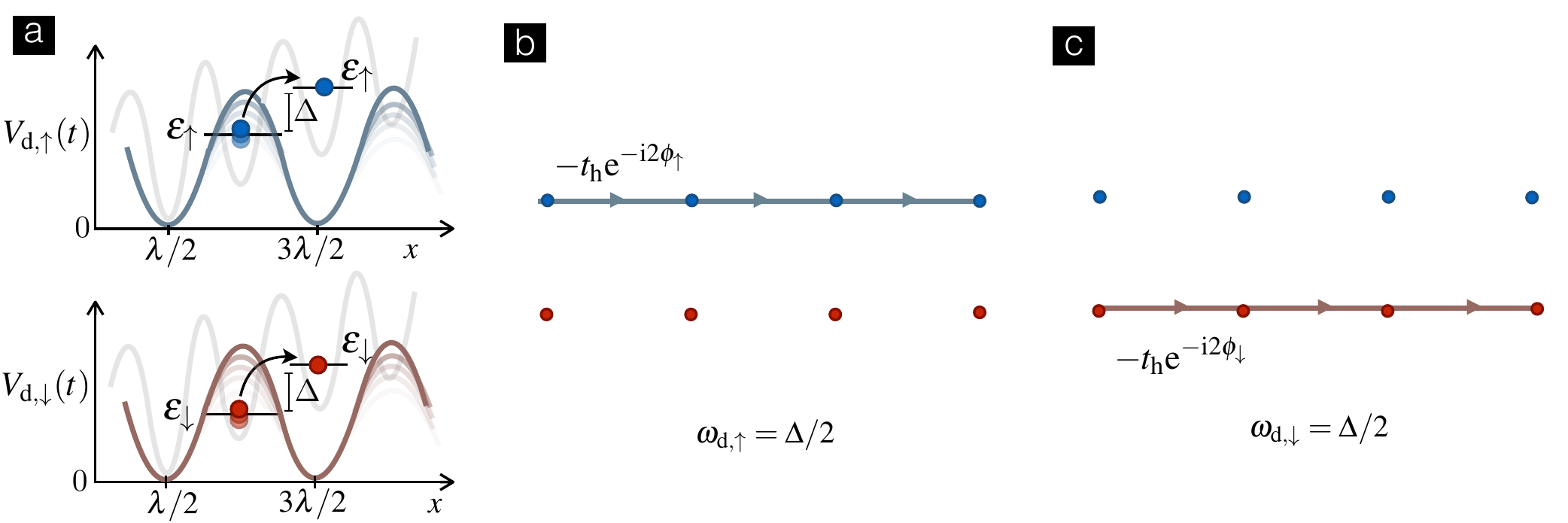}
\caption{ {\bf Photon-assisted tunnelling scheme for a binary hyperfine mixture of ultracold atoms in an optical lattice: } {\bf (a)} Atoms in a given hyperfine state can tunnel to the neighbouring potential well preserving its hyperfine state by absorbing energy from an additional optical lattice whose intensity is periodically modulated. {\bf (b)} The intra-leg tunnelling of $\ket{{\uparrow}}$ (blue circles) is activated by the weak modulated lattice with $\omega_{\rm d,\uparrow}=\Delta/2$, such that the tunnelling picks the modulation phase $\ee^{-\ii2\phi_{\uparrow}}$. {\bf (c)} The intra-leg tunnelling of $\ket{{\downarrow}}$ (red circles) is activated by the weak modulated lattice with $\omega_{\rm d,\downarrow}=\Delta/2$, such that the tunnelling picks the modulation phase $\ee^{-\ii2\phi_{\downarrow}}$.}
\label{fig_driving_scheme}
\end{figure*}

\subsubsection{Intensity-modulated optical lattice}

 To obtain the complex horizontal hopping in Fig.~\ref{fig_ladder_scheme}{\bf (c)}, we must re-activate the intra-leg tunnelling inhibited by the lattice tilting~\eqref{eq:tilt}, and dress it with the correct Peierls phases. This can be achieved by introducing a periodic driving provided by an additional optical lattice with a modulated intensity (see Fig.~\ref{fig_driving_scheme}{\bf (a)}), namely
\beq
\label{eq:driving}
H_{\rm driving}=\sum_{i,\sigma} V_{\rm d,\sigma}(t)\sin^2( k_{\rm d}x_i^0)f_{i,\sigma}^{{\dagger}}f_{i,\sigma}^{\phantom{\dagger}}, 
\eeq
where the intensity of the ac-Stark shifts for each of the hyperfine states is modulated periodically in time, $V_{\rm d,\sigma}(t)=V_{\rm d,0}\sin(\omega_{\rm d,\sigma}t-\phi_{\sigma})$ with frequency $\omega_{\rm d,\sigma}$ and a phase $\phi_{\sigma}$.
In analogy with the lattice tilting~\eqref{eq:tilt}, we have assumed that the amplitude fulfills $V_{\rm d,0}\ll V_{0x}$, such that we can neglect any periodic modulation on the bare tunnelling of the original Hubbard model~\eqref{eq:tU}, and consider instead a periodic driving of the on-site energies~\eqref{eq:driving}. More importantly, since this lattice must be very shallow, one can minimise the scattering of photons from the excited state in the calculation of the two-photon ac-Stark shift, which can lead to the aforementioned problematic heating effects in the opposite regime of deep state-dependent optical lattices~\cite{gd_gauge_fields}. As an interesting alternative, we note that  a time-dependent magnetic-field gradient~\cite{gradient_modulation} can also be exploited to obtain state-dependent time-periodic modulations.

\subsubsection{Effective Creutz-Hubbard Hamiltonian}

Once the ingredients have been introduced, let us show how the synthetic Creutz-Hubbard model is obtained. If the wavelength of this intensity-modulated lattice is twice that of the original lattice $\lambda_{\rm d}=2\lambda$, only the even sites will be subjected to the periodic modulation (see Fig.~\ref{fig_driving_scheme}{\bf (a)}). Moreover, if the resonance condition is met, namely $n\omega_{\rm d,\sigma}=\Delta$ for $n\in\mathbb{Z}$, the nearest-neighbour tunnelling can be restored by absorbing energy quanta from the periodic drive. By setting $\omega_{\rm d,\sigma}=\Delta/2$, the intra-leg tunnelings acquire the phase of the corresponding drivings (see Fig.~\ref{fig_driving_scheme}{\bf (b)}-{\bf (c)}), and lead to
\beq
H_{\rm h}=-t\frak{J}_{2}\left(\frac{V_{\rm d,0}}{\Delta}\right)\sum_{i,\sigma}\ee^{-\ii 2\phi_{\sigma}}f_{i+1,\sigma}^{{\dagger}}f_{i,\sigma}^{\phantom{\dagger}}+{\rm H.c.},
\eeq
where $\frak{J}_{2}(x)$ is the second-order Bessel function of the first class. Let us note that the phases of the periodic modulations must fulfil $\phi_{\uparrow}=-\phi_{\downarrow}$, and that the Raman-assisted tunnelling~\eqref{Raman} will get off-resonantly modified by this driving, leading to the final inter-leg tunnelling
\beq
H_{\rm d}= \frac{\Omega}{2}\frak{J}_{0}\left(\frac{V_{\rm d,0}}{\Delta}\right)\sum_{i}\left(f_{i+1,\downarrow}^{\dagger} f_{i,\uparrow}^{\phantom{\dagger}}+f_{i+1,\uparrow}^{\dagger} f_{i,\downarrow}^{\phantom{\dagger}}\right) +{\rm H.c.},
\eeq
where $\frak{J}_{0}(x)$ is the zero-order Bessel function. We finally note that the on-site Raman transitions, which were previously neglected under a rotating-wave approximation $|\Omega_{12}|,|\Omega_{13}|\ll\Delta$, can also become activated due to the intensity-modulated lattice. This contributes with an spurious vertical inter-leg tunnelling 
\beq 
H_{\rm v}=\frac{\Omega_{12}}{2}\frak{J}_{2}\left(\frac{2V_{\rm d,0}}{\Delta}\sin\left(\half(\phi_{\uparrow}-\phi_{\uparrow})\right)\right)\sum_i f_{2i,\downarrow}^{\dagger} f^{\phantom{\dagger}}_{2i,\uparrow}+{\rm H.c.}.
\eeq
 At this point we should stress that the above derivation does not pose any constraint on the ratio $V_{\rm d,0}/\Delta$. Hence, we can set this ratio such that we hit a zero of the corresponding Bessel function, e.g. $2V_{\rm d,0}\sin\left(\frac{\phi_{\uparrow}-\phi_{\uparrow}}{2}\right)=5.1356\Delta$, which yields a coherent destruction of the spurious tunnelling against the energy offset~\cite{cdt_ol}, such that we can make $H_{\rm v}\approx 0$.

All the ingredients discussed so far combine to give us the desired laser-assisted tunnelling that implements the Creutz-ladder kinetic Hamiltonian $H_{\rm C}=H_{\rm h}+H_{\rm d}$~\eqref{or_cl}. This becomes clear after the mapping
\beq
\label{eq:synth_dim}
f_{i,\uparrow}^{\phantom{\dagger}}\to c_{i,\rm u}^{\phantom{\dagger}},\hspace{2ex} f_{i,\downarrow}^{\phantom{\dagger}}\to c_{i,\rm d}^{\phantom{\dagger}},
\eeq
and the associated identification of the Creutz-ladder parameters with the microscopic cold-atom ones
\beq
\label{eq:tunn_parameters}
\begin{split}
t_{\rm h}={+}t&\frak{J}_{2}\left(\frac{V_{\rm d,0}}{\Delta}\right),\hspace{2ex}\theta=2\phi_{\uparrow},\hspace{2ex}t_{\rm d}=-\frac{\Omega}{2}\frak{J}_{0}\left(\frac{V_{\rm d,0}}{\Delta}\right).
 \end{split}
\eeq
In addition, if the Raman beams in Eq.~\eqref{Raman} are slightly detuned from the resonance, we would have an additional term 
\beq
\label{eq:imb_parameters}
H_{\rm local}=\frac{\delta}{2}\sum_i\left(f_{i,\uparrow}^\dagger f_{i,\uparrow}^{\phantom{\dagger}}-f_{i,\downarrow}^\dagger f_{i,\downarrow}^{\phantom{\dagger}}\right),
\eeq where $\delta$ is the Raman detuning, and we work in a rotating frame. By identifying
\beq
	\Delta\epsilon=\frac{\delta}{2},
\eeq
 we obtain the last ingredient of the kinetic energy, namely the leg imbalance in Eq.~\eqref{eq:creutz-hubbard_pert}.

So far, we have only been concerned with the kinetic term of the Creutz ladder. As studied for other type of periodic drivings~\cite{per_mod_tunneling,3_body} or~\cite{gauge_ol,int_dep_pat}, the Hubbard interactions present in the microscopic model~\eqref{eq:hubbard} may have an important impact in the dressed tunnelings if they modify the resonance conditions. In this work, we assume that such resonances are avoided, which permits mapping the s-wave scattering of the cold atoms~\eqref{eq:hubbard} onto the required Hubbard interaction of the Creutz ladder~\eqref{eq:H_H} directly:
\beq
\label{eq:int_parameters}
V_{\rm v}=U_{\uparrow\downarrow},\hspace{2ex} V_{\rm h}=0.
\eeq 

We have thus finished with the derivation of the synthetic Creutz-Hubbard model in the cold-atom setup. To summarise, one can explore the properties of the imbalanced Creutz-Hubbard Hamiltonian in Eq.~\eqref{eq:pi_ch} by applying a laser-assisted-tunnelling scheme to a two-component gas of fermionic atoms loaded 1D optical lattice, such that the components play the role of the two legs of the ladder~\eqref{eq:synth_dim}. The Hamiltonian parameters can be controlled experimentally using the mappings in Eqs.~\eqref{eq:tunn_parameters},~\eqref{eq:imb_parameters}, and~\eqref{eq:int_parameters}, together with the expressions~\eqref{eq:tU} and~\eqref{eq:OmegaR}, as a guiding principle. The particular Hamiltonian~\eqref{eq:pi_ch} with $\tilde{t}=t\frak{J}_0(V_{\rm d,0}/\Delta)$, $\Delta\epsilon=\delta/2$, and $V_{\rm v}=U_{\uparrow\downarrow}$, is obtained by tuning the ratio of the bare and Raman tunnelings $t/\Omega=-2\frak{J}_0(V_{\rm d,0}/\Delta)/\frak{J}_2(V_{\rm d,0}/\Delta)$, such that the strength of the inter- and intra-leg tunnelings is equal, and by fixing the phases such that $\phi_{\uparrow}=\pi/4=-\phi_{\downarrow}$ leads to a net $\pi$-flux.

\section{Conclusions and Outlook}
\label{sec:conclusions}

In this work, we have advanced on the understanding of the competition between topological features and interaction effects in quantum many-body systems. By focussing on the paradigmatic Creutz-Hubbard ladder,
we have developed a variety of analytical tools, and  performed a thorough numerical analysis, unveiling characteristic features and general methods that could be applied to other strongly-correlated topological insulators/superconductors. Moreover,
our predictions can be readily tested in table-top experiments with ultracold fermionic atoms in optical lattices, which may serve as quantum simulators of interacting topological phases.

By using a pair of electronic states of the atoms as synthetic legs of the ladder, 
and by exploiting laser-assisted tunnelling to control their dynamics,
we have shown that the physics of a variant of the Creutz ladder can be implemented in a one-dimensional optical lattice. Such a variant is obtained by applying a 
Zeeman splitting between the two legs of the ladder, which puts the system into the AIII class of topological insulators. To the best of our knowledge, topological insulators in such a symmetry class have not been  realized in cold-atom experiments yet. Moreover, by
adding on-site repulsive interactions only, we can induce phase transitions of different universality classes into an orbital ferro- or paramagnetic phase.
Surprisingly, we found that the expected Dirac CFT transition line at weak interactions splits into two Majorana CFT critical lines, once the Hubbard term dominates over the Zeeman term: these findings are also numerically supported by the scaling of entanglement entropy.
In addition, we provided an understanding of these topological phase transitions in the language of quantum impurity physics,
shedding new light on the hybridisation mechanism of the edge states.
Finally, we have identified experimentally accessible signatures to test our theoretical predictions.

From the above results, we are convinced that the synthetic Creutz-Hubbard ladder can become a workhorse in the theoretical and experimental study of correlated topological insulators, as it allows for a very neat understanding of the underlying phase diagram, and a clear path to implement the model with ultracold atoms.
This could be helpful in order to gain even deeper insight into the tri-critical point,  serving as a guide to construct an effective quantum field theory that describes the mechanism underlying the splitting of the central charge into the two critical lines. It would be very interesting to study the imbalanced Creutz-Hubbard model at different fillings, and to explore the possibility of finding topological phases of matter  that disappear for vanishing interactions. In this respect, the analytic and numerical methods hereby presented may be generalised to other fillings, allowing to go beyond  mean-field arguments that support the existence of such interesting ground states.
  
\textbf{Acknowledgments.--} We thank L. Mazza and M. Burrello for critical reading of the paper before submission, L. Tarruell for interesting discussions on the experimental details, and L. Tagliacozzo and J. Stasinska for fruitful discussions.

J.J. thanks Studienstiftung des deutschen Volkes for financial support. 
M.R. thanks the KITP Santa Barbara for hospitality within the Visiting Program ``Synthetic Quantum Matter'', during which part of the manuscript writing was performed. 
A.B. acknowledges support from Spanish MINECO Projects FIS2015-70856-P,  and CAM PRICYT project  QUITEMAD+ S2013/ICE-2801.
A.P., S.R. and M.L. acknowledge financial support from Fundaci\'o Cellex, from the European Union (ERC-2013-AdG Grant No. 339106 OSYRIS, FP7-ICT-2011-9 No. 600645 SIQS, H2020-FETPROACT-2014 No. 641122 QUIC, FP7/2007-2013 Grant No. 323714 Equam), from Spanish MINECO Project (FIS2013-46768-P FOQUS, SEV-2015-0522 Severo Ochoa), from the Generalitat de Catalunya (2014 SGR 874) and from CERCA Programme / Generalitat de Catalunya.
Some of the MPS simulations were run by J.J. and M.R. on the Mogon cluster of the JGU (made available by the CSM and AHRP), with a code based on a flexible Abelian Symmetric Tensor Networks Library, developed in collaboration with the group of S. Montangero at the University of Ulm.



\begin{thebibliography}{100}
\bibitem{comment_edge_bulk}
				The energies of the bulk bands for the finite chain can differ from those of the periodic chain by an intensive quantity related to the missing bond $t_{\rm imb}$ connecting the edges. In the thermodynamic limit, $N\to\infty$, this difference will be irrelevant for any extensive observable, or any quantity that depends on a sum over all possible bulk energies.
\end{thebibliography}
\end{document}